\documentclass[acmsmall]{acmart}
\usepackage{algorithm}
\usepackage[noend]{algpseudocode}
\usepackage{algorithmicx}
\usepackage{enumitem}
\usepackage{multirow}
\usepackage{subfig}
\newtheorem{theorem}{Theorem}[section]

\newtheorem{observation}{Observation}[section]

\definecolor{modified}{rgb}{1,0,0}
\AtBeginDocument{%
  \providecommand\BibTeX{{%
    \normalfont B\kern-0.5em{\scshape i\kern-0.25em b}\kern-0.8em\TeX}}}

\setcopyright{acmlicensed}
\acmJournal{PACMMOD}
\acmYear{2023} \acmVolume{1} \acmNumber{1} \acmArticle{61} \acmMonth{5} \acmPrice{15.00}\acmDOI{10.1145/3588915}

\begin{document}

\title{Towards Generating Hop-constrained s-t Simple Path Graphs}


\author{Yuzheng Cai}
\email{yuzhengcai21@m.fudan.edu.cn}
\affiliation{
  \institution{School of Data Science, Fudan University}
 \city{Shanghai}
  \country{China}
}

\author{Siyuan Liu}
\email{liusiyuan19@fudan.edu.cn}
\affiliation{
  \institution{School of Data Science, Fudan University}
 \city{Shanghai}
  \country{China}
}

\author{Weiguo Zheng}
\email{zhengweiguo@fudan.edu.cn}
\affiliation{
  \institution{School of Data Science, Fudan University}
 \city{Shanghai}
  \country{China}
}

\author{Xuemin Lin}
\affiliation{
  \institution{Antai College of Economics and Management, Shanghai Jiao Tong University}
  \city{Shanghai}
  \country{China}
}
\email{lxue@cse.unsw.edu.au}









\begin{abstract}
Graphs have been widely used in real-world applications, in which investigating relations between vertices is an important task. In this paper, we study the problem of generating the $k$-hop-constrained $s$-$t$ simple path graph, i.e., the subgraph consisting of all simple paths from vertex $s$ to vertex $t$ of length no larger than $k$.
To our best knowledge, we are the first to formalize this problem and prove its NP-hardness on directed graphs. 
To tackle this challenging problem, we propose an efficient algorithm named \textit{EVE}, which exploits the paradigm of edge-wise examination rather than exhaustively enumerating all paths. 
Powered by essential vertices appearing in all simple paths between vertex pairs, \textit{EVE} distinguishes the edges that are definitely (or not) contained in the desired simple path graph, producing a tight upper-bound graph in the time cost $\mathcal{O}(k^2|E|)$. 
Each remaining undetermined edge is further verified to deliver the exact answer.
Extensive experiments are conducted on $15$ real networks. The results show that \textit{EVE} significantly outperforms all baselines by several orders of magnitude. Moreover, by taking \textit{EVE} as a built-in block, state-of-the-art for hop-constrained simple path enumeration can be accelerated by up to an order of magnitude.
\end{abstract}

\ccsdesc[500]{Information systems~Database management system engines}

\keywords{simple path graph, essential vertices, upper-bound graph}

\received{July 2022}
\received[revised]{October 2022}
\received[accepted]{November 2022}

\maketitle

\section{Introduction}
In graph analytics and applications, mining relations between two given vertices is one of the fundamental problems, which helps to make use of connections from a vertex $s$ to another vertex $t$, or to investigate influences or similarities between them based on graph topology \cite{JOIN2020, JOIN2021, ESTI, PathEnum, QbS, ShortestWidestRouting, ConstrainedShortestPath}.
In this paper, we focus on the problem of $k$-hop-constrained $s$-$t$ simple path graph generation. 
In many real-life applications, relations between vertices $s$ and $t$ can be captured by enumerating all $s$-$t$ simple paths, where each path is no longer than a user-specific hop constraint $k$ \cite{JOIN2020, JOIN2021}. Let us consider the following example.

\subsection{Motivation}
\label{sec: motivation}

\begin{figure}[t]
  \centering
  \includegraphics[width=0.76\linewidth]{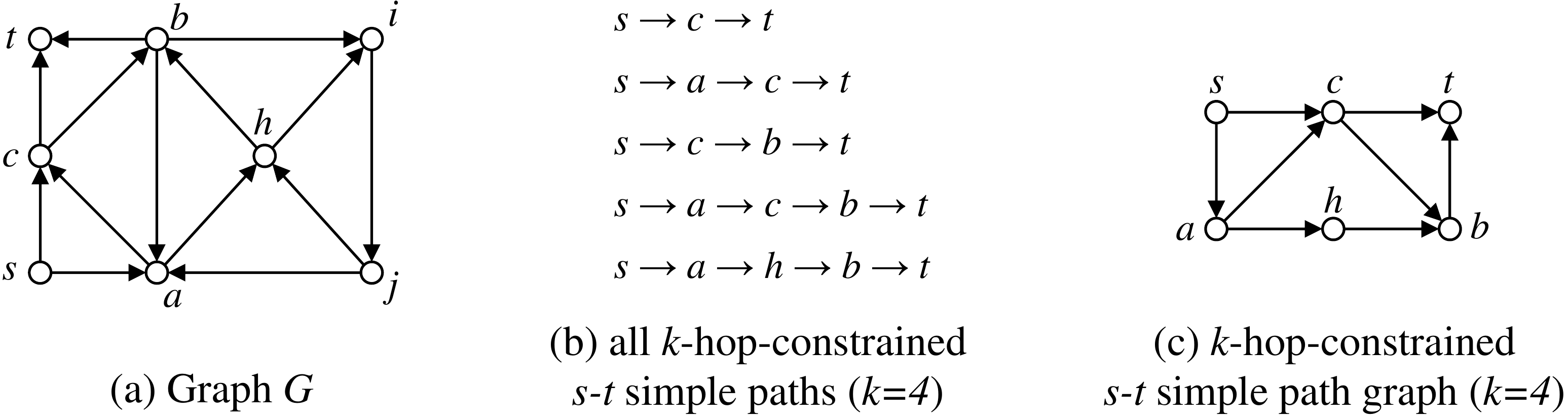}
  \caption{Motivation example}
  \label{fig:introduction}
  \vspace{-0.05in}
\end{figure}

For graph $G$ in Figure \ref{fig:introduction}(a), all simple paths from vertex $s$ to $t$ with hop constraint $k=4$ are presented in Figure \ref{fig:introduction}(b).
The problem of hop-constrained $s$-$t$ simple path enumeration has been well studied,
and various novel techniques have been proposed \cite{TDFS, JOIN2020, JOIN2021, PathEnum}.
However, a user may be overwhelmed by the huge number of paths listed. There may be many common vertices and edges in these hop-constrained $s$-$t$ simple paths for large strongly cohesive communities \cite{JOIN2021}, as illustrated in Figure \ref{fig:introduction}(b). 
Such overlaps motivate us to generate a \textit{k-hop-constrained $s$-$t$ Simple Path Graph} (denoted as $SPG_k(s,t)$), a subgraph of input graph $G$, each edge of which is contained in at least a simple path from $s$ to $t$ not longer than $k$.
Figure \ref{fig:introduction}(c) is an $s$-$t$ simple path graph with $k=4$ for graph $G$ in Figure \ref{fig:introduction}(a). It captures the main structure of connections between $s$ and $t$, while preserving concision by avoiding repeated vertices and edges in all paths listed in Figure \ref{fig:introduction}(b).

\noindent\textbf{Applications.}
The problem of $k$-hop-constrained $s$-$t$ simple path graph generation has a wide range of applications, e.g.,  fraud detection, relation visualization, and accelerating other algorithms.

 \textit{\underline{Fraud Detection}.}
In financial systems, transaction activities can be modeled as a directed graph, where each vertex represents a person or account, and each edge $e(u,v)$ represents a transaction from $u$ to $v$.
A simple cycle in such a graph is a strong indication of fraudulent activity or even a financial crime like money laundering \cite{Alibaba, JOIN2020, JOIN2021, PathEnum}.
For a certain transaction $e(t,s)$, by extracting vertices and edges in all simple cycles containing $e(t,s)$, all fraudsters and fraudulent transactions involved can be identified.
In practice, the maximum length is specified for the desired cycles \cite{PathEnum}.
Clearly, generating the hop-constrained simple path graph from $s$ to $t$ will immediately produce the target fraudsters and transactions.
Similar to previous works \cite{Alibaba, JOIN2020, JOIN2021, PathEnum}, non-simple cycles are not considered here, since they may contain other cycles not related to the current edge $e(t,s)$ (i.e., not participating current fraud). Involving them may impose unnecessary repeated punishments and increase downstream workloads (e.g., monitoring and investigation).

\begin{figure}[t]
	\centering
	\subfloat[5-hop simple path graph from Consumer 114898 to 578 on the who-trust-whom online social network \cite{Signed-Network}]{\includegraphics[width=.5\columnwidth]{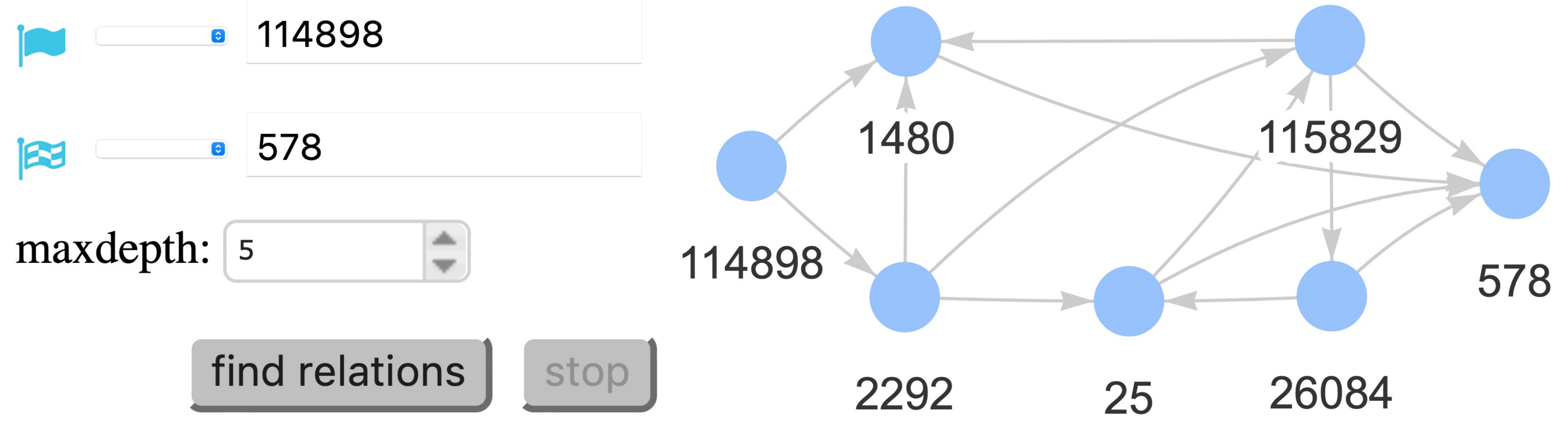}}\hspace{10pt}
	\subfloat[Numbers of edges in $SPG_k$ for graphs \textit{wn} and \textit{uk}]{\includegraphics[width=.44\columnwidth]{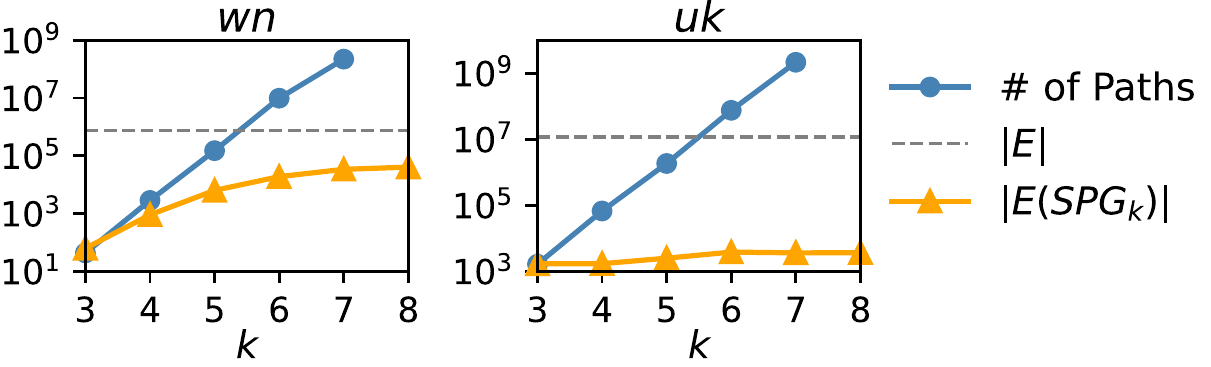}}
	\caption{Illustrations for relation visualization and the growing numbers of edges w.r.t. $k$}
    \label{fig:relfinder-pathcount}
\end{figure}

 \textit{\underline{Relation Visualization}.} 
To discover non-obvious relationships between two entities for decision making and risk reduction \cite{RelFinder2010}, visualizing hop-constrained $s$-$t$ simple path graphs is widely desired in visualization systems, supporting various types of data, such as biological linked data, scientific datasets, and enterprise information networks \cite{RelFinder2009,RelFinder2010,SciVis,BioVis,JOIN2021}. 
For example, given two user-specified nodes $(s, t)$ and some constraints (e.g., path length), RelFinder \cite{RelFinder2009, RelFinder2010} executes SPARQL queries to extract constrained $s$-$t$ simple paths one by one, then displays the $s$-$t$ simple path graph (see Figure~\ref{fig:relfinder-pathcount}(a)) instead of listing all paths (e.g., Figure \ref{fig:introduction}(b)).
Directly generating the simple path graph not only avoids costly path enumeration, but also provides an initial result which can be further processed for advanced options if required. 

Besides the applications above, $k$-hop-constrained $s$-$t$ simple path graph generation can be also used to accelerate other graph algorithms that take simple path graph generation as a built-in block, like (i) hop-constrained simple path enumeration \cite{TDFS, JOIN2020, JOIN2021, PathEnum}, (ii) quality of service (QoS) routing \cite{routing}, and (iii) minimum length-bounded $s$-$t$-cut problem \cite{MinCut}. 
They also take vertex pair $(s,t)$ and hop constraint $k$ as input, and edges out of $SPG_k(s,t)$ will not be considered for output.

\subsection{Challenges and Contributions}
\label{sec:Challenges and Contributions}

\textbf{Challenges.}
Since a huge search space may be involved when searching from $s$ to $t$ \cite{JOIN2020, JOIN2021}, generating $k$-hop-constrained $s$-$t$ simple path graph $SPG_k(s,t)$ is computationally expensive.
A straightforward solution is to enumerate all $k$-hop-constrained $s$-$t$ simple paths, then put all edges and vertices of these paths together to obtain $SPG_k(s,t)$, suffering from the time cost $\mathcal{O}(\delta k|E|)$ where $\delta$ is the number of paths and $|E|$ is the number of edges in $G$ \cite{JOIN2020, JOIN2021, PathEnum}. 
Since $\delta$ may grow exponentially w.r.t. the number of hops $k$ in real graphs \cite{JOIN2020, JOIN2021}, enumerating all paths is far from efficient given the truth that the number of edges in the desired simple path graph is bounded by $|E|$.
Take Figure~\ref{fig:relfinder-pathcount}(b) as an example which presents the number of edges in $SPG_k(s,t)$ and the number of $s$-$t$ simple paths by varying $k$ from 3 to 8 on two graphs from  NetworkRepository \cite{NetworkRepository} (the results are averaged on 1000 random queries for each $k$).

To enhance the efficiency, we need to address three questions.

\begin{itemize}[leftmargin=*]
    \item Considering the expensive cost of enumerating all the  $k$-hop-constrained $s$-$t$ simple paths, is it possible to generate $SPG_k(s,t)$ without enumerating all paths? 
    \item Since the problem of generating $SPG_k(s,t)$  is NP-hard (as proved in Section \ref{sec:problem-definition}), can we obtain approximate results with high quality in polynomial time? 
    \item 
    As it may need to generate some paths for obtaining the exact $SPG_k(s,t)$, how can we accelerate searching process in practice?
\end{itemize}

To answer these questions, a method named \textit{\underline{E}ssential \underline{V}ertices based \underline{E}xamination} (shorted as \textit{EVE}) is proposed for building the $k$-hop-constrained $s$-$t$ simple path graph $SPG_k(s,t)$.
Instead of enumerating all paths, \textit{EVE} examines whether each edge is involved in $SPG_k(s,t)$ by introducing \textit{essential vertices} that summarise common vertices appearing in paths between certain vertex pairs (see Section~\ref{sec: Essential Vertex}).
Powered by \textit{essential vertices} with a much smaller cost compared to the naive solution above, most failing edges can be filtered out in $\mathcal{O}(k^2|E|)$ without enumerating any paths.

To find the essential vertices efficiently, a propagating computation paradigm is developed, in which a forward-looking pruning strategy is exploited to reduce computational costs. 
Benefiting from essential vertices, each edge can be easily assigned a label indicating whether the edge is definitely (or not) contained in $SPG_k(s,t)$, otherwise promising but not verified (called undetermined edges). 
Thus, an upper-bound graph  $SPG^u_k(s,t)$ (see Section~\ref{sec: upperbound}) can be acquired.
Finally, each undetermined edge will be verified by finding a valid simple path passing the edge.
Carefully designed search orders are employed to further accelerate the verification.

\begin{table}
  \caption{Summary of notations}
  \footnotesize
  \label{tab:notations}
  \begin{tabular}{ll}\hline 
    \textbf{Notation} & \textbf{Description} \\ \hline
    $G$, $G^r$ & A directed graph and a reversed graph of $G$\\
    $d_{max}$, $d_{avg}$ & Maximum and average degree of $G$\\
    $s$, $t$ & Source and target vertex for query\\
    $\Delta (s,t)$ & Shortest distance from $s$ to $t$\\
    $p(s,t), p^\ast(s,t)$ & A path and simple path from $s$ to $t$\\
    $V(p), E(p), |p|$ & Vertex set, edge set, and length of $p(s,t)$\\
    $P_k(s,t)$ & All $k$-hop-constrained $s$-$t$ paths\\
    $P_k^*(s,t)$ & All $k$-hop-constrained $s$-$t$ simple paths\\
    $EV_k(s,t)$ & {Essential vertex set of $P_k(s,t)$}\\ 
    $EV_k^*(s,t)$ & {Essential vertex set of $P_k^*(s,t)$}\\ 
    $SPG_k(s,t)$, $SPG_k$ & $k$-hop-constrained $s$-$t$ simple path graph\\
    $SPG^u_k(s,t)$, $SPG^u_k$ & Upper-bound graph of $SPG_k$\\  \hline
\end{tabular}
\end{table}

\noindent\textbf{Contributions.}
Our contributions are summarised as follows.

\begin{itemize}[leftmargin=*]
    \item To the best of our knowledge, we are the first to formalize the problem of $k$-hop-constrained $s$-$t$ simple path graph generation, which is motivated by a wide range of applications. We also prove that this problem is NP-hard for directed graphs.
    \item To address this challenging problem, we develop the paradigm of examining whether an edge is involved in a hop-constrained $s$-$t$ simple path graph powered by \textit{essential vertices}, instead of exhaustively enumerating all paths. 
    \item We propose an efficient approach, namely \textit{EVE}, consisting of three components, including propagation for \textit{essential vertices}, upper-bound graph computation, and verification. 
    The first two steps deliver an upper-bound graph in $\mathcal{O}(k^2|E|)$ time. 
    Each undetermined edge is verified with carefully designed search orders.
    \item We conduct comprehensive experiments on real graphs to compare \textit{EVE} against baseline methods. 
    \textit{EVE} significantly outperforms the baselines in terms of time efficiency by up to 4 orders of magnitude. 
    Experimental results also demonstrate the tightness of the computed upper-bound graph which contains less than $0.05\%$ redundant edges for most graphs. 
    Furthermore, \textit{PathEnum} \cite{PathEnum} (state-of-the-art for hop-constrained $s$-$t$ simple path enumeration) can be accelerated by up to an order of magnitude powered by the proposed simple path graph.
\end{itemize}

\section{Problem Definition and Overview}
\label{sec:Overview} 
We formally define the problem of $k$-hop-constrained $s$-$t$ simple path graph generation, and then give an overview of the proposed \textit{EVE}.
Table~\ref{tab:notations} lists the frequently-used terms throughout the paper.

\subsection{Problem Definition}
\label{sec:problem-definition}
Let $G=(V, E)$ denote a directed graph, where $V$ is a vertex set and $E\subseteq V \times V$ is an edge set. Let $e(u, v)\in E$ represent a directed edge from vertex $u$ to $v$. 
The number of vertices and edges are denoted as $|V|$ and $|E|$, respectively. 
The maximum and average vertex degree are denoted as $d_{max}$ and $d_{avg}$, respectively.
Reversing direction of all edges in $G$ leads to a reversed graph, denoted by $G^r = (V, E^r )$.

Given the source vertex $s$ and target vertex $t$, an $s$-$t$ path $p(s,t)=\{s=v_0, v_1, \ldots, v_{l-1}, v_l=t\}$ denotes a directed vertex sequence from $s$ to $t$, where $(v_{i-1}, v_i) \in E$. The vertex set and edge set of path $p(s,t)$ are denoted as $V(p)$ and $E(p)$, and its length is $|p|=l$. 
A simple path $p^\ast(s,t)$ is such a path without duplicate vertices, i.e., $\forall v_i , v_j \in p^\ast(s,t)$ s.t. $0 \leq i < j \leq l$, $v_i \ne v_j$. 
All paths from $s$ to $t$ with length $l \leq k$ is denoted as a set $P_k(s,t)=\{p_1(s,t),$ $p_2(s,t),\ldots,p_n(s,t)\}$ where $|p_i(s,t)| \leq k$.  
Similarly, $P_k^*(s,t)$ denotes all $s$-$t$ simple paths with length $l \leq k$.

\begin{definition} 
($k$-hop-constrained $s$-$t$ Simple Path Graph). Given a graph $G$ and a query $\langle s,t,k \rangle$, the $s$-$t$ simple path graph with hop constraint $k$, denoted as $SPG_k(s,t)$=($V^\ast,E^\ast$), is a subgraph of $G$  such that $V^\ast = \cup_{p^\ast \in P_k^\ast(s,t)}{V(p^\ast)}$ and $E^\ast = \cup_{p^\ast \in P_k^\ast(s,t)}{E(p^\ast)}$.
\end{definition}

For ease of presentation, $SPG_k(s,t)$ can be simplified as $SPG_k$. 
For each edge $e \in E^\ast$, there must exist a $k$-hop-constrained $s$-$t$ simple path passing through $e$.

\begin{example}
For graph $G$ in Figure~\ref{fig:introduction}(a), when the hop constraint $k=4$, all $k$-hop-constrained $s$-$t$ simple paths $P_k^\ast(s,t)$ are shown in Figure~\ref{fig:introduction}(b), while $SPG_k(s,t)$ is shown in Figure~\ref{fig:introduction}(c).
\end{example}

\textbf{Problem Statement.}  (\textit{$k$-hop-constrained $s$-$t$ Simple Path Graph Generation}).
For a directed graph $G$, given a query  $\langle s,t,k \rangle$, where $s$ and $t$ are two vertices in $G$ ($s \neq t$) and $k$ is the hop constraint, the task is to find the hop-constrained $s$-$t$ simple path graph $SPG_k(s,t)$.

\subsection{Hardness Analysis}
\label{sec:hardness}

Next, we prove that the problem of generating $SPG_k(s,t)$ is NP-hard by introducing \textit{Fixed Subgraph Homeomorphism Problem} \cite{ProofNP}.
Given a pattern graph $T$ and an input directed graph $G$, 
\textit{a homeomorphism mapping $f$} consists of node-mapping $f^v$ and edge-mapping $f^e$. $f^v$ maps vertices of $T$ to vertices of $G$ and $f^e$ maps edges of $T$ to simple paths in $G$, where the involved simple paths in $G$ must be pairwise node-disjoint but allowing shared start and end nodes.

\begin{definition}
(Fixed Subgraph Homeomorphism Problem, FSH for short) \cite{ProofNP}.
For a fixed pattern graph $T$, given an input directed graph $G$ with node-mapping $f^v$ from $T$ to $G$ specified, does $G$ contain a subgraph $G^*$ homeomorphic to $T$ ? 
\end{definition}

\begin{figure}[t]
  \centering
  \includegraphics[width=.95\linewidth]{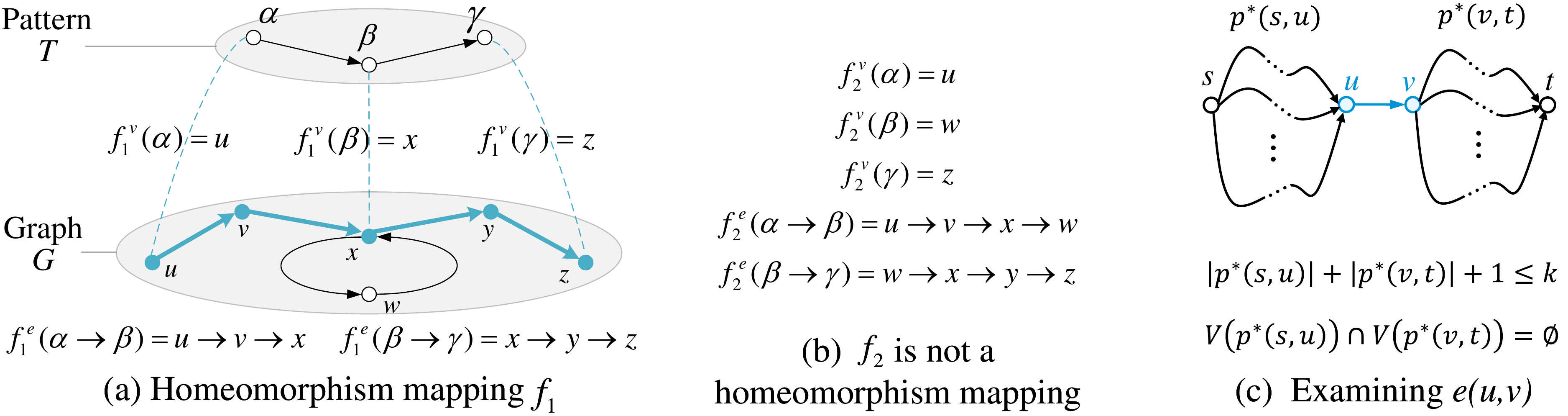}
  \caption{Illustration for fixed subgraph homeomorphism problem and Oberservation~\ref{obs:edge check}}
  \label{fig:proofNP-obs}
\end{figure}

\begin{example}
For the fixed pattern graph $T$ of three vertices and two edges: $\alpha \rightarrow \beta \rightarrow \gamma$, given a directed graph $G$ with a node-mapping specified as $f^v_1(\alpha)=u$, $f^v_1(\beta)=x$, and $f^v_1(\gamma)=z$ in Figure~\ref{fig:proofNP-obs}(a), the answer to FSH problem is $yes$. 
It is because there exists a subgraph $G^*$ (bold vertices and edges) and $f_1$ consisting of $f^v_1$ and $f^e_1$ is a homeomorphism mapping.

However, when node-mapping is specified as $f^v_2(\alpha)=u$, $f^v_2(\beta)=w$, and $f^v_2(\gamma)=z$,  the answer is $no$. As shown in Figure~\ref{fig:proofNP-obs}(b), $f^e_2$ is not a homeomorphism edge-mapping, since path $u \rightarrow v \rightarrow x \rightarrow w$ and $w \rightarrow x \rightarrow y \rightarrow z$ share the common vertex $x$.
\end{example}

When the pattern graph $T$ is fixed to be a path with two edges connecting three distinct vertices (i.e., $\alpha \rightarrow \beta \rightarrow \gamma$), it can be proved that determining whether $T$ is subgraph homeomorphic to $G$ under a node mapping $f^v$ is NP-complete \cite{ProofNP}. 

\begin{theorem}
\label{the: NP}
The $k$-hop-constrained $s$-$t$ simple path graph generation problem for directed graphs is NP-hard.
\end{theorem}

\begin{proof}
For the fixed pattern graph $T$: $\alpha \rightarrow \beta \rightarrow \gamma$, the FSH problem can be reduced to \textit{k-hop-constrained $s$-$t$ simple path graph generation} in $\mathcal{O}(|V|^2)$ time. 
Specifically, given a directed graph $G$ with a node-mapping $f^v(\alpha)=s$, $f^v(\beta)=r$, and $f^v(\gamma)=t$ where $s$, $r$, and $t$ are three distinct vertices in $G$, the FSH problem can be solved by generating $SPG_k(s,t)$ for $k=1,2,\cdots,|V|-1$.

(1) If $\exists k$ s.t. $r \in V(SPG_k(s,t))$, there exists a simple path $p^\ast$ from $s$ to $t$ through $r$. $p^\ast$ can be split into node-disjoint simple paths $p^\ast(s,r)$ and $p^\ast(r,t)$. Hence, the answer to the FSH problem is $yes$.

(2) If $\forall k$, $r \not\in V(SPG_k(s,t))$, there does not exist node-disjoint simple paths $p^\ast(s,r)$ and $p^\ast(r,t)$. Otherwise, joining $p^\ast(s,r)$ and $p^\ast(r,t)$ together will obtain a simple path $p^\ast(s,t)$ passing through $r$. Hence, the answer to the FSH problem is $no$.

Thus, if the $k$-hop-constrained $s$-$t$ simple path graph generation can be solved in polynomial time, the FSH problem will be solved in polynomial time too, which contradicts its NP-completeness. 
\end{proof}

We also discuss the hardness of generating $SPG_k(s,t)$ when $k$ is much smaller than $|V|$. 

\begin{theorem}
\label{the: n^c NP}
For any constant $\epsilon \in (0,1)$, the problem of generating $SPG_k(s,t)$ given $k=|V|^\epsilon$ on directed graphs is NP-hard.
\end{theorem}

\begin{proof}
For each graph $G(V,E)$ of the FSH problem, we construct a graph $G'(V',E')$ by adding $(|V|^\frac{1}{\epsilon}-|V|)$ isolated vertices to $G$ such that $|V'|=|V|^\frac{1}{\epsilon}$ and $E'=E$.
Similar to the proof of Theorem~\ref{the: NP}, we can generate $SPG_k(s,t)$ over graph $G'$ with $k=|V'|^\epsilon$ and return $yes$ for the FSH problem iff $r\in V(SPG_k(s,t))$.
Since adding $(|V|^\frac{1}{\epsilon}-|V|)$ isolated vertices costs polynomial time, if we can generate $SPG_k(s,t)$ over $G'$ given $k=|V'|^\epsilon$ in polynomial time, the FSH problem will be solved in polynomial time too, which contradicts the NP-completeness of the FSH problem. 
\end{proof}

\begin{theorem}
\label{the:FPT}
The problem of $k$-hop-constrained $s$-$t$ simple path graph generation is ﬁxed-parameter tractable (FPT).
\end{theorem}

\begin{proof}
Let us consider the
\textit{Directed k-(s, t)-Path Problem} that decides whether there is an $s$-$t$ simple path of length $k$ in a directed graph $G$ \cite{directeds-t}. 
\textit{Directed k-(s, t)-Path Problem} is ﬁxed-parameter tractable and can be solved in $\mathcal{O}(2^{\mathcal{O}(k)}E)$ \cite{color-coding} as a generalization of $k$-path, i.e., it can be solved in polynomial time for $k=\log |V|$.

The solution to \textit{Directed k-(s, t)-Path Problem} can be also used to build $SPG_k(s,t)$.
Specifically, $SPG_k(s,t)$ can be obtained by checking each edge $e(u,v)$. It holds that $e(u,v) \in SPG_k(s,t) \Leftrightarrow \exists$ a simple path $p^\ast$ through $e(u,v)$ s.t. $|p^\ast| \leq k$.
For each edge $e(u,v)$, we construct an auxiliary graph $G'(V',E')$ by inserting a new vertex into every edge in $G$ (i.e., splitting each edge of $G$ into two edges connected by a new vertex) except $e(u,v)$, resulting in $|V'|=|V|+|E|-1$ and $|E'|=2|E|-1$. 
Next, we invoke the solver of \textit{Directed $k'$-(s, t)-Path Problem} for $k'=1,3,5,\cdots,2k-1$, where $k'$ is an odd number.
Then we have: $\exists k'$ s.t. there exists an $s$-$t$ simple path $q^\ast$ of length $k'$ in $G'$ $\Leftrightarrow \; \exists$ an $s$-$t$ simple path $p^\ast$ of length $l=\frac{k'+1}{2} \leq k$ passing through $e(u,v)$ in $G$.
Proof is trivial by noticing that each of the other edges in $G$ is split into two edges in $G'$, and any $s$-$t$ simple path of odd length in $G'$ must pass through $e(u,v)$.

The reduction above takes the $\mathcal{O}(k|E|^2)$ time. Since \textit{Directed k-(s, t)-Path Problem} is FPT, generating $SPG_k(s,t)$ is also FPT.
\end{proof}

Notice that though the FPT algorithm mentioned in the proof above is theoretically feasible, it has a significant failure rate as shown in extensive experiments \cite{PathWay}. 

\subsection{Overview of Our Approach}
\label{sec:General Ideas}

Enumerating all $k$-hop-constrained $s$-$t$ simple paths with naive DFS, leading to the time cost $\mathcal{O}(|V|^k)$. 
It is still computationally expensive even equipped with state-of-the-art algorithms for path enumeration.
To efficiently generate $SPG_k(s,t)$, we focus on the question: Given an edge $e(u,v)$, does it belong to $SPG_k(s,t)$?

\begin{observation}
\label{obs:edge check}
$e(u,v) \in SPG_k(s,t) \Leftrightarrow \exists \; p^\ast(s,u), p^\ast(v,t)$ s.t. (1) $|p^\ast(s,u)| + |p^\ast(v,t)| + 1 \leq k$ and (2) $V(p^\ast(s,u)) \cap V(p^\ast(v,t))=\emptyset$.
\end{observation}

As illustrated in Figure \ref{fig:proofNP-obs}(c), Observation~\ref{obs:edge check} holds since any desired path $p^\ast(s,t)$ through $e(u,v)$ can be decomposed into $p^\ast(s,u)$, $e(u,v)$, and $p^\ast(v,t)$ s.t. conditions (1) and (2) are satisfied.
By iterating each edge $e(u,v)$ and each pair $( p^\ast(s,u), p^\ast(v,t) )$ for checking the conditions (1) and (2), we can find out all edges of $SPG_k$.
However, computing and iterating all pairs of $p^\ast(s,u)$ and $p^\ast(v,t)$ is still unacceptable in both time and space cost. 

\begin{figure}[t]
  \centering
  \includegraphics[width=\linewidth]{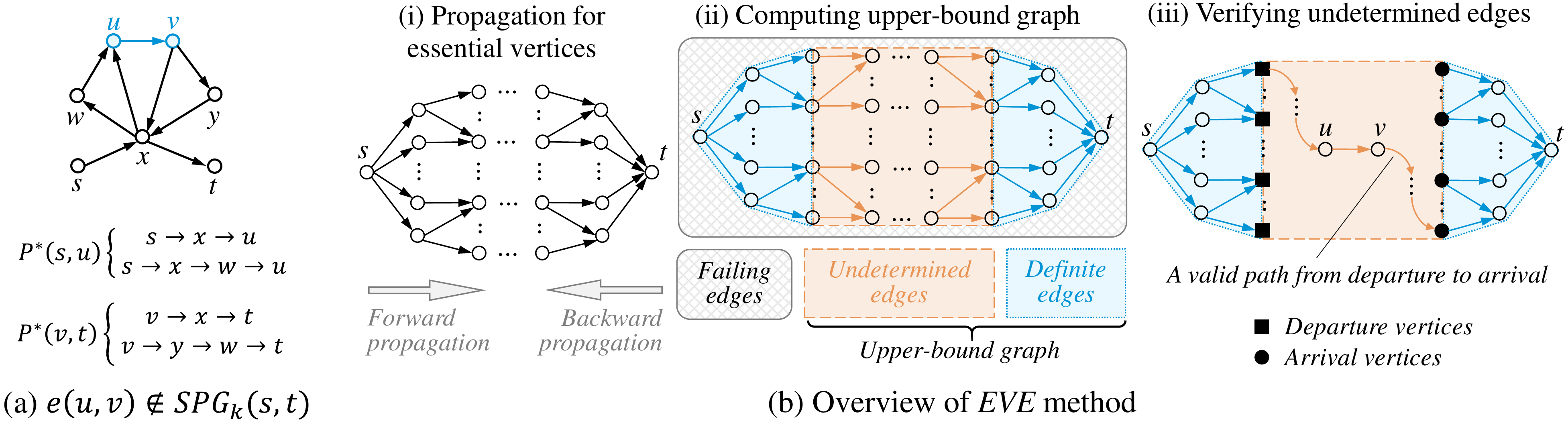}
  \caption{Illustration for general ideas of the \textit{EVE} method}
  \label{fig:overview}
  \vspace{-0.1in}
\end{figure}

Intuitively, by ensuring shortest distances $\Delta(s,u)+\Delta(v,t)+1 \leq k$, it is easy to determine whether an edge $e(u,v)$ satisfies condition (1).
However, there are still many failing edges $e(u,v)$ caused by some vertices appearing in both $p^\ast(s,u)$ and $p^\ast(v,t)$ (i.e., do not satisfy condition (2)).
Thus, we introduce the concept of \textit{Essential Vertices} (formally defined in Definition~\ref{def:essential vertices}) for vertices appearing in all $k$-hop-constrained simple paths $p^\ast(s,u)$ (or $p^\ast(v,t)$), denoted by $EV^\ast_k(s,u)$ (or $EV^\ast_k(v,t)$).
Take edge $e(u,v)$ in Figure~\ref{fig:overview}(a) as an example, $EV^\ast_k(s,u) = EV^\ast_k(v,t) = \{x\}$ when $k=6$ since vertex $x$ lies on all $p^\ast(s,u)$ and $p^\ast(v,t)$. Thus,  $EV^\ast_k(s,u) \cap EV^\ast_k(v,t) \neq \emptyset$, concluding that $e(u,v) \not\in SPG_k(s,t)$.
In this paper, we prove that edge $e(u,v) \not\in SPG_k(s,t)$ if $EV_{k_f}^*(s,u)\cap EV_{k_b}^*(v,t) \neq \emptyset$ for $\forall k_f, k_b$ s.t. $k_f+k_b+1 \leq k$ (see Theorem~\ref{the:upper bound}), which helps to effectively identify failing edges in practice.
Experiments show that when $k=6$, among edges only satisfying condition (1), there are about $30\%$ redundant edges not in $SPG_k$ (averaged on 1000 random queries for all datasets in Table~\ref{tab:datasets}).
After filtering with essential vertices, there are only about $0.1\%$ redundant edges left.

We summarise our proposed method \textit{EVE} in three phases which avoids the expensive cost of iterating all pairs of $p^\ast(s,u)$ and $p^\ast(v,t)$ for examining each edge $e(u,v)$, as shown in Figure~\ref{fig:overview}(b).

\noindent \textbf{(1) Propagation for Essential Vertices.} 
To obtain $EV^\ast_k(s,w)$ and $EV^\ast_k(w,t)$ for any vertex $w$, we conduct forward propagation from $s$ and backward propagation from $t$, where essential vertices are propagated layer by layer.
To further reduce unnecessary propagating cost, we develop forward-looking pruning strategy that takes the shortest distance from $s$ to $u$ (resp. $\Delta (s,u)$) and the shortest distance from $v$ to $t$ (resp. $\Delta (v,t)$) into consideration, ensuring that $EV^\ast_l(s,w)$ and $EV^\ast_l(w,t)$ computed in the current propagation step $l$ will be used.
Adaptive bi-directional search is exploited to efficiently obtain such shortest distances before propagation.

\noindent \textbf{(2) Computing upper-bound graph.}
All edges are categorized into three sets including failing edges, undetermined edges, and definite edges.
Specifically, failing edges are found and filtered out based on essential vertices by Theorem~\ref{the:upper bound}, while an upper-bound graph $SPG_k^u$ can be obtained from the remaining edges (see Section~\ref{sec: upper bound edge}).
Furthermore, each edge in $SPG_k^u$ will be distinguished as a definite edge or an undetermined edge (see Section~\ref{sec: answer edge}).

\noindent \textbf{(3) Verifying Undetermined Edges.} 
\textit{Departures} and \textit{arrivals} (formally defined in Section~\ref{sec:departures and arrivals}) are  computed on $SPG_k^u(s,t)$, which are boundary vertices connecting undetermined and definite edges.
For each undetermined edge $e(u,v)$, DFS-oriented search is conducted to find a valid simple path from \textit{departure} to \textit{arrival} through $e(u,v)$ by Theorem~\ref{the: verification requirements}. 
If such valid path exists, $e(u,v) \in SPG_k(s,t)$. 
With carefully designed searching orders, verification process is further accelerated.

When $k\leq 4$, the upper-bound graph $SPG^u(s,t)$ computed above always equals $SPG_k(s,t)$ (Theorem~\ref{cor:query4}), indicating no false-positive edges will be produced. 
When $k \leq 5$, it only takes $\mathcal{O}(|E|)$ time and space for generating the exact $SPG_k(s,t)$ (Theorem~\ref{the:time complexity k<=5}).

\section{Essential Vertex Computation}
\label{sec: Essential Vertex} 

We first present the principle of introducing essential vertices (Section~\ref{subsec:essentical vertex}). To compute essential vertices, a propagation based algorithm is proposed (Section~\ref{sec:Lower Bound}). To speed up the computation, we propose a forward-looking pruning strategy (Section~\ref{sec: adaptive bi-directional BFS}).

\subsection{Essential Vertices}
\label{subsec:essentical vertex}

As discussed in Section~\ref{sec:General Ideas}, Observation~\ref{obs:edge check} abstracts the existence of a path $p^\ast(s,t)$ through $e(u,v)$ by checking the constraints for each edge individually.  
This insight has the advantage of substantially avoiding edges' repetitive visits, but validating each edge still demands significant overhead.
Thus, greedy or pruning methods should be introduced to minimize edge-wise verification. 

\textbf{Pruning Principle.} 
When $P_{k-1}^\ast(s,u) \neq \emptyset$ and $P_{k-1}^\ast(v,t) \neq \emptyset$, edge $e(u,v)$ cannot be contained in $SPG_k(s,t)$ iff:
\underline{$\forall p^\ast(s,u) \in P_{k-1}^\ast(s,u)$, $p^\ast(v,t) \in P_{k-1}^\ast(v,t)$ s.t. $|p^\ast(s,u)|+|p^\ast(v,t)|< k$, $V(p^\ast(s,u)) \cap $}
\underline{$V(p^\ast(v,t))\neq \emptyset$.}
In other words, though it is possible to form a path of length $|p^\ast(s,t)|\leq k$, duplicate vertices are inevitable in each of such paths.
Thus, it provides opportunities for identifying the case 
$e(u,v)\notin SPG_k(s,t)$ when $P_{k-1}^\ast(s,u)$ and $P_{k-1}^\ast(v,t)$ satisfy certain features.
We extract the key features associated with the vertices, named as \textbf{essential vertices}.

\begin{definition} (Essential Vertices).
\label{def:essential vertices}
Given query vertices $s$ and $t$ in $G$, essential vertices for $P_{l}^\ast(s,u)$ are denoted as $EV_l^*(s,u)$, i.e., the set of vertices that are contained in all simple paths not passing through $t$ in $P_l^*(s,u)$. Formally, we have
\begin{equation}
\label{EVS}
EV_l^\ast(s, u)= \bigcap\limits_{p_i^\ast\in P_l^\ast(s,u) \; s.t. \; t \not\in V(p_i^\ast)}^{}V(p_i^\ast).
\end{equation}
\end{definition}

The vertices in $EV_l^*(s, u)$ are essential, since to form a simple path $p^\ast(s,t)$, there will be no $p^\ast(s,u)$ with $\lvert p^\ast(s,u) \rvert \leq l$ if all vertices in $EV_l^*(s, u)$ are removed.
Similarly, we have
\begin{equation}
\label{EVS vt}
EV_l^\ast(v, t)= \bigcap\limits_{p_i^\ast\in P_l^\ast(v,t) \; s.t. \; s \not\in V(p_i^\ast)}^{}V(p_i^\ast).
\end{equation}

\begin{example}
For graph $G$ in Figure~\ref{fig:introduction}(a), essential vertices $EV_l^\ast(s, \cdot)$ and $EV_l^\ast(\cdot, t)$ for $1\leq l \leq 6$ are shown in Figure~\ref{fig:EV}(a)-(b).
Take vertex $b$ as an example, $P_2^\ast (s,b) = \{ \{s,c,b\} \}$, $P_3^\ast (s,b) = \{ \{s,c,b\},\{s,a,h,b\},$ $\{s,a,c,b\} \}$.
Thus, $EV_2^\ast (s,b) = \{s,c,b\}$ and $EV_3^\ast (s,b) = \{s,b\}$.
\end{example}

Based on $EV_l^*(s, u)$ and $EV_l^\ast(v, t)$, we can determine which edges can be definitely excluded from $SPG_k(s,t)$ by Theorem~\ref{the:upper bound}.

\begin{lemma}
\label{lemma:deflated edge check}
If  $e(u,v)\in SPG_k(s,t)$, it holds that $\exists \; EV^\ast_{k_f}(s,u)$ and $\exists \ EV^\ast_{k_b}(v,t)$ s.t. (1) $k_f + 1 + k_b \leq k$, and (2) $EV_{k_f}^*(s,u)\cap EV_{k_b}^*(v,t)=\emptyset$; but not necessarily the reverse.
\end{lemma}

\begin{proof}
\textbf{Sufficiency}. 
As $e(u,v) \in SPG_k(s,t)$, there must exist a simple path $p^*(s,u)$ whose length is $k_f$ and $p^*(v,t)$ whose length is $k_b$ subject to the condition that $k_f+k_b+1 \leq k$ and $V(p^\ast(s,u))\cap V(p^\ast(v,t))=\emptyset$. 
Meanwhile, $EV^\ast_{k_f}(s,u)\subseteq V(p^*(s,u))$ and $EV^\ast_{k_b}(v,t)\subseteq V(p^*(v,t))$. Therefore, $EV^\ast_{k_f}(s,u) \cap EV^\ast_{k_b}(v,t)=\emptyset$.\\
\textbf{Necessity}. 
Consider edge $e(b,a)$ as a counterexample in Figure~\ref{fig:introduction}(a). 
If $k=7$, for $k_f=3 \land k_b=2$ satisfying condition (1), $EV_3^\ast (s,b) = \{s,b\}$ and $EV_2^\ast (a,t) = \{a,c,t\}$.
Thus, $EV_3^\ast (s,b) \cap EV_2^\ast (a,t) = \emptyset$ and condition (2) also holds.
However, no simple path from $s$ to $t$ passing $e(b,a)$ can be formed. Thus, the necessity is not established.
\end{proof}

\begin{theorem}
\label{the:upper bound}
If $\forall 0\leq k_f,k_b \leq k-1$ s.t. $k_f+1+k_b \leq k$, we have
(1) $EV_{k_f}^\ast(s,u)$ or $EV_{k_b}^\ast(v,t)$ does not exist (i.e., there is no simple path $p^\ast(s,u)$ or $p^\ast(v,t)$ of length no larger than $k_f$ or $k_b$, respectively); or (2)  $EV_{k_f}^\ast(s,u)\cap EV_{k_b}^\ast(v,t)\neq \emptyset$, then $e(u,v)\notin SPG_k(s,t)$.
\end{theorem}

The proof of Theorem~\ref{the:upper bound} is omitted since it is the contrapositive statement of {Lemma~\ref{lemma:deflated edge check}.}
Hence, we can prune as many edges as possible to achieve a practically tight upper bound of the desired $k$-hop-constrained $s$-$t$ simple path graph (for more details please refer to Section~\ref{sec: upperbound}).
Thus, overall time cost can be greatly reduced. 

\subsection{Propagating Computation}
\label{sec:Lower Bound}

\begin{figure}[t]
  \centering
  \includegraphics[width=0.8\linewidth]{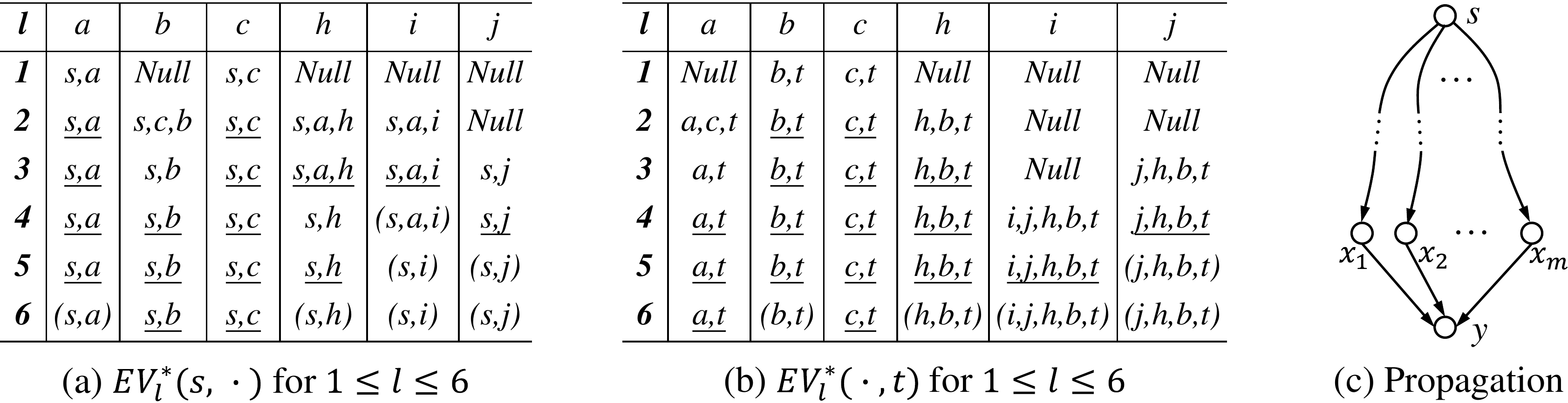}
  \caption{Illustrations for essential vertex computation}
  \label{fig:EV}
\end{figure}

Directly computing essential vertices based on the definition is challenging. Specifically, to compute $EV^\ast_l(s,u)$, all simple paths $p^\ast(s,u)$ with  $\lvert p^\ast(s,u) \rvert \le l$ have to be enumerated, and their vertices need to be intersected. 
To address the problem, we propose an efficient propagating computation for essential vertices, which works on the fact that \textit{intersecting all simple paths' vertices is equivalent to intersecting all paths' vertices (the correctness is guaranteed by Theorem~\ref{the:path = simple path}}). 
Similar to  $EV^*_l(s,u)$, let $EV_l(s,u)$ denote the intersection of vertices of all paths not passing through $t$ in $P_l(s,u)$.
Formally,
\begin{equation}
\label{EV}
EV_l(s, u)= \bigcap\limits_{p_j\in P_l(s,u) \; s.t. \; t \not\in V(p_j)}^{}V(p_j).
\end{equation}

The basic principle is that the essential vertices of $EV_{l}(s,y)$ can be  computed based on the essential vertices of $y$'s incoming neighbors.
Note that for ensuring $t \not\in V(p_j)$ in Equation~(\ref{EV}), vertex $t$ will never be visited when computing $EV_l(s,y)$.
Let $In(y)$ denote the incoming neighbors of vertex $y$.
Formally, the propagating computation follows the recursive formula below.
\begin{equation}
\label{EV update}
    EV_{l}(s,y)=\bigcap_{x_i \in In(y)} \left( EV_{l-1}(s,x_i) \cup \{ y \} \right)
\end{equation}

Intuitions of Equation~(\ref{EV update}): 
Any path $p(s,y)$ with length $|p(s,y)|\leq l$ is composed of two parts, i.e., a path $p(s,x_i)$ and an edge $e(x_i,y)$, where $x_i \in In(y)$.  
Taking Figure~\ref{fig:EV}(c) as an example, suppose $x_1$, $x_2$, $\ldots$, $x_m$ are all the in-neighbors of $y$ s.t. $P_{l-1}(s,x_i) \neq \emptyset$ ($1\leq i\leq m$). 
A path $p(s,y)$ with $\lvert p(s,y) \rvert \leq l$ must pass through an edge $e(x_i,y)$. 
Given essential vertices $EV_{l-1}(s,x_i)$ for paths $p(s, x_i)$ with length $|p(s, x_i)| \leq l-1$, all the essential vertices from $s$ to $y$ through $e(x_i,y)$ within length $l$ are $EV_{l-1}(s,x_i)\cup \{y\}$. Thus, intersecting essential vertices through $e(x_i,y)$ for all $x_i \in In(y)$ leads to $EV_{l}(s,y)$.

Propagation is terminated when $l > k-1$, since only $EV^\ast_l(s,u)$ for $l\leq k-1$ are required in Theorem~\ref{the:upper bound}.
To propagate essential vertices through edges within ($k-1$)-hop edges from $s$, we introduce an algorithm \textit{Forward Propagation} (see Algorithm~\ref{Forward_Propagation}). 
It first initializes all essential vertex sets in line~\ref{line: initializaton}, then stimulates the process of path generation from $s$ and maintains a queue called $Frontier$.
For a certain length $l$, the algorithm visits out-neighbors $y \in Out(x)$ for vertices $x$ in the $Frontier$, and intersects $EV_{l}(s,y)$ with $EV_{l-1}(s,x) \cup \{y\}$ in lines~\ref{line: intersect for y start}-\ref{line: intersect for y end}. 
From another prospective, for each vertex $y$ explored in layer $l$, the algorithm propagates the set of essential vertices from its in-neighbors by Equation~(\ref{EV update}).
Note that in line~\ref{line:inherit}, if vertex $u$ has not been visited in current layer $l$, 
$EV_l(s,u)$ can inherit vertices from $EV_{l-1}(s,u)$, since paths within $l-1$ hops are  not longer than $l$.

\begin{algorithm}[t] 
	\caption{Forward Propagation}
	\label{Forward_Propagation}
	\footnotesize
	\begin{algorithmic}[1]
		\Require Hop constraint $k$, source vertex $s$, and graph $G=(V,E)$.
		\Ensure $ EV_{l}(s,y)$ for $ y\in V$ and $1\leq l< k$.
		\State{$EV_{l}(s,u) \leftarrow null$  for each $u \in V$ where $l=0$ to $k-1$ } \label{line: initializaton}
		\State $Frontier \leftarrow \{ s\}$; $EV_0(s,s) \leftarrow \{ s \}$
		\For{$l=1$ to $k-1$}
		    \State{$nextFrontier \leftarrow \emptyset$}
		    \For{$x$ in $Frontier$}
		        \For{$y$ in $Out(x)$ s.t. $y\neq s$ $\land$ $y\neq t$} \label{line: intersect for y start}
        	        \If{$ EV_{l}(s,y)=null$} 
    		            \State $EV_{l}(s,y) \leftarrow EV_{l-1}(s,x) \cup \{y\} $
    		        \Else
		                \State $EV_{l}(s,y) \leftarrow EV_{l}(s,y)\cap ( EV_{l-1}(s,x) \cup \{y\})$
		            \EndIf \label{line: intersect for y end}
    		        \State Add $y$ to $nextFrontier$ \label{line: add to frontier}
		        \EndFor
		    \EndFor
		    \State $EV_{l}(s,u) \leftarrow EV_{l-1}(s,u)$ for each $u \in V$ if $EV_{l}(s,u)=null$ \label{line:inherit}
		    \State{$Frontier\leftarrow nextFrontier$}
		\EndFor 
	\end{algorithmic} 
\end{algorithm}

To compute essential vertices $EV_l^\ast(v,t)$ of any length $l$ from every vertex $v$ to $t$, 
we run this algorithm on the reversed graph $G^r$, taking $t$ as the source vertex.
This process is called \textit{Backward Propagation}.
Next, we show that Algorithm~\ref{Forward_Propagation} is correct as $EV_l(s,u)=EV_l^*(s,u)$.

\begin{theorem}
\label{the:path = simple path}
$\forall$ vertex pair $(s, u)$ and $l\geq 1$, $EV_l(s, u) = EV_l^\ast(s, u)$.
\end{theorem}

\begin{proof}
When $P_l^\ast(s,u) \subset P_l(s,u)$, we have $P_l(s,u) \neq \emptyset$ and there exists a path $p(s,u)\in P_l(s,u)$ reusing vertices. 
We can always find a repeatedly used vertex $x$ that taking out the first $x$ in the path (denoted as $x_1$) and the last one (denoted as $x_n$), such that the path  $p(s,u)=\{s,\ldots, $ $x_1,\ldots,x_n, \ldots,u \}$ split by $x_1$ and $x_n$ follows the constraint that no vertex $y$ exists both in $p(s,x_1)$ and $p(x_n,u)$. 
If we combine $p(s,x_1)$ with $p(x_n,u)$, the new path is a path $p^x(s,u)$ from $s$ to $u$ with no vertex reused, whose length $|p^x(s,u)|<|p(s,u)| \leq l$. 
As $V(p^x(s,u)) \subseteq V(p(s,u))$, it holds that $\forall p \in P_l(s,u)$, there must exist a simple path $p^\ast \in P_l^\ast(s,u) \subset P_l(s,u)$ satisfying $V(p^\ast)\cap V(p)=V(p^\ast)$. 
Thus, $EV_l(s, u)= \bigcap_{p_j\in P_l(s,u) \; s.t. \; t \not\in V(p_j)}V(p_j)=\bigcap_{p_i^\ast\in P_l^\ast(s,u) \; s.t. \; t \not\in V(p_i^\ast)}V(p_i^\ast)=EV_l^\ast(s, u)$.
\end{proof}

\subsection{Forward-looking Pruning Strategy}
\label{sec: adaptive bi-directional BFS}

In forward propagation, essential vertices are computed for every vertex $u$ satisfying $\Delta(s,u)<k$.
However, not all $EV_l^\ast (s,u)$ will be useful for filtering unpromising edges.
Intuitively, for any vertex $u$ s.t. $\Delta(s,u)+\Delta(u,t)>k$, computing $EV^\ast_l(s,u)$ is meaningless due to length constraint.
Moreover, even for vertex $u$ satisfying $\Delta(s,u)+\Delta(u,t)\leq k$, unnecessary computation can also be avoided.

\begin{theorem}
\label{the: forward looking}
If $l+\Delta(u,t)>k$, computing $EV_l^\ast(s,u)$ contributes nothing to concluding any edge $e(u,v) \not\in SPG_k(s,t)$ by Theorem~\ref{the:upper bound}.
\end{theorem}

\begin{proof}
For edge $e(u,v)$, if $l+\Delta(u,t)>k$, there does not exist $EV^\ast_{k_b}(v,t)$ s.t. $l + 1 + k_b \leq k$.
Thus, statement (1) in Theorem~\ref{the:upper bound} holds when $k_f=l$.
If such $EV_l^\ast (s,u)$ are not computed (i.e., not exist), statement (1) still holds when $k_f=l$.
\end{proof}

Based on Theorem~\ref{the: forward looking}, we develop a forward-looking pruning strategy which concerns the steps needed for reaching $t$ in future.
Specifically, $l+\Delta(y,t) \leq k$ is required when propagating from vertex $x$ to $y$ at $l^{th}$ step.
In other words, line~\ref{line: intersect for y start} of Algorithm~\ref{Forward_Propagation} is changed to:
\textit{for $y$ in $Out(x)$ s.t. $y\neq s$ $\land$ $y\neq t$ $\land$ $l+\Delta(y,t) \leq k$}.
Note that when $l+\Delta(y,t) > k$, $y$ will not be added to $nextFrontier$ in line~\ref{line: add to frontier} for exploring its out-neighbors.
It is because $y$ has no descendants $y'$ at any step $l'$ ($l'>l$) satisfying $l'+\Delta(y',t)\leq k$,
otherwise $l+\Delta(y,t) \leq l+(l'-l+\Delta(y',t)) \leq k$.

\begin{example}
Given graph $G$ (Figure~\ref{fig:introduction}(a)) and $k=7$, 
those essential vertices in parenthesis in Figure~\ref{fig:EV}(a) will not be computed.
For example, $EV^\ast_l(s,i)$ for $l>3$ can be omitted since $\Delta(i,t)=4$.
\end{example}

In practice, when $EV^\ast_l(s,u)$ are the same for different $l$, we only store the first one since the others can refer to it.
For instance, those underlined essential vertices in Figure~\ref{fig:EV}(a) can be omitted.
Similarly, time and space cost of backward propagation can also be reduced.
Note that shortest distances $\Delta(s,y)$ and $\Delta(y,t)$ should be computed before starting propagation.
Intuitively, we can conduct forward BFS from $s$ and backward BFS from $t$, but better solutions exist.

We can only compute distances for vertices $y$ satisfying $\Delta(s,y)+\Delta(y,t) \leq k$ and assign $+\infty$ for the other vertices $w$,
since $l+\Delta(w,t) \ge \Delta(s,w)+\Delta(w,t) >k$ will stop forward propagation for $w$ at step $l$.
As shown in Figure~\ref{fig:searchspace-edgelabel}(a), bi-directional BFS \cite{bidirectional, QbS} explores forward from $s$ and backward from $t$ with equal depth, then continues forward (resp. backward) search for the remaining steps over edges explored backward (resp. forward).
To improve the performance, we can use \textit{Adaptive Bi-directional Search} \cite{AdaptiveBidirectional_1, AdaptiveBidirectional_2}, in which forward and backward BFS  start simultaneously.
It adaptively explores in the direction having a smaller frontier at each step, until the total depth reaches $k$.

\textbf{Time Complexity of Computing Essential Vertices.} To compute essential vertices, 
adaptive bi-directional search first obtains shortest distances in $\mathcal{O}(\lvert E \rvert)$ time. 
Then Algorithm~\ref{Forward_Propagation} initializes $EV_l(s,u)$ for all vertices $u$  with $0\leq l < k$. 
For each $l$, it updates $EV_l(s,y)$ for all out-neighbors $y$ of vertices $x$ in the frontier by intersecting $EV_{l-1}(s,x) \cup \{ y \}$ with $EV_{l}(s,y)$, which needs at most $\lvert E \rvert$ intersections since there are $\lvert E \rvert$ edges in the graph. 
In each intersection, each essential vertices set contains at most $l+1$ vertices.
Thus, total time complexity is $\mathcal{O}( k^2 \lvert E \rvert )$.

\textbf{Space Complexity of Computing Essential Vertices.} Adaptive bi-directional search maintains two BFS frontiers requiring $\mathcal{O}(\lvert V \rvert)$ complexity.
Essential vertices sets from $s$ and to $t$ are maintained for each vertex of all length $l$ ($0 \leq l< k$). 
For each length, up to  $l$ essential vertices exist. 
Thus, total space complexity is $\mathcal{O}(k^2 \lvert V \rvert)$. 

\section{Essential Vertex based Upper Bound}
\label{sec: upperbound}

As shown in step (ii) of the overview (Figure~\ref{fig:overview}(b)), we divide all edges in $G$ into three groups:
\textit{Failing edges}, which are definitely not in $SPG_k(s,t)$, labeled as ``0'';
\textit{Undetermined edges}, which are a promising edge in $SPG_k(s,t)$ but not verified, labeled as ``1'';
\textit{Definite edges}, which are definitely contained in $SPG_k(s,t)$, labeled as ``2''.
Section~\ref{sec: upper bound edge} focuses on ruling out failing edges to form an upper-bound graph,
while definite edges are identified in Section~\ref{sec: answer edge}.

\subsection{Computing the Upper-Bound Graph}
\label{sec: upper bound edge}

\begin{figure}[t]
  \centering
  \includegraphics[width=\linewidth]{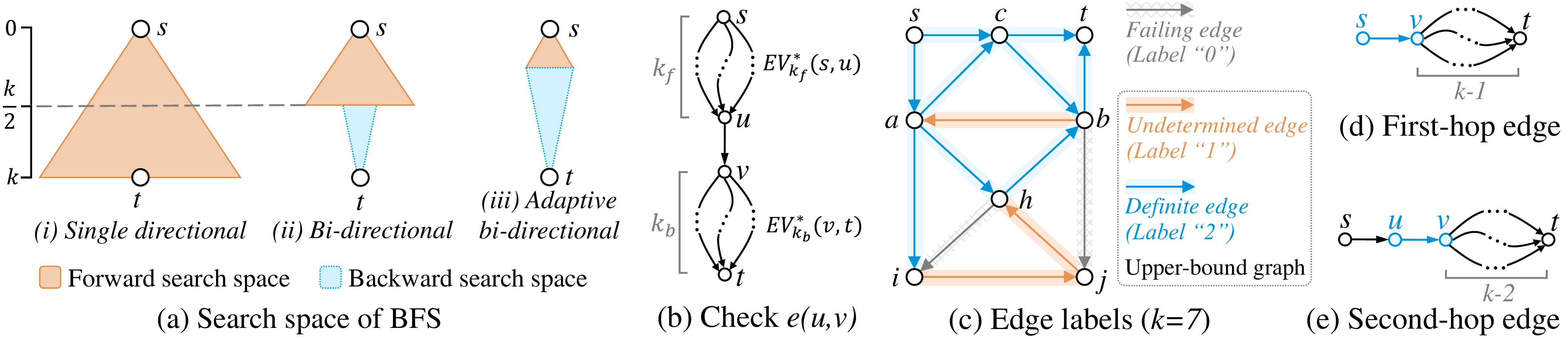}
  \caption{Illustrations for BFS search space and edge labeling}
  \label{fig:searchspace-edgelabel}
\end{figure}

As proved in Theorem~\ref{the: NP}, generating the $k$-hop-constrained $s$-$t$ simple path graph is an NP-hard problem. 
Therefore, it motivates us to find an upper bound of the target path graph to narrow down the search space for computing $SPG_k(s,t)$. 

As illustrated in Figure~\ref{fig:searchspace-edgelabel}(b), according to Theorem~\ref{the:upper bound}, for each edge $e(u,v)$ we can iterate all pairs of integers ($k_f$, $k_b$) s.t. $k_f+1+k_b \leq k$. 
We can conclude that $e(u,v) \not\in SPG_k(s,t)$ if none of the pairs ($k_f$, $k_b$) satisfies the constraints that (1) $EV_{k_f}^\ast(s,u)$ and $EV_{k_b}^\ast(v,t)$ exist; and (2) $EV_{k_f}^\ast(s,u)\cap EV_{k_b}^\ast(v,t)= \emptyset$.
Otherwise, it contributes to forming an upper-bound graph of $SPG_k(s,t)$.

\begin{definition} 
\label{def:upper bound graph}
(Upper-bound graph of $SPG_k(s,t)$). An upper-bound graph of $SPG_k(s,t)$, denoted as $SPG_k^u(s,t)=(V^u,E^u)$, is a subgraph of $G$ {such that: $ e(u,v) \in E^u \Leftrightarrow$} $\exists \; EV^\ast_{k_f}(s,u)$ and $\exists \ EV^\ast_{k_b}(v,t)$ s.t. (1) $k_f + 1 + k_b \leq k$, and (2) $EV_{k_f}^*(s,u)\cap EV_{k_b}^*(v,t)=\emptyset$.
\end{definition}

To form $SPG_k^u(s,t)$, we iterate each edge $e(u,v)$ to check whether it is contained in $SPG_k^u(s,t)$ based on Definition~\ref{def:upper bound graph}.
Note that the edges out of adaptive bi-directional search space (Section~\ref{sec: adaptive bi-directional BFS}) are not considered, since they cannot satisfy distance constraint.

\begin{example}
Given the graph $G$ in Figure~\ref{fig:introduction}(a) and $k=7$, the upper-bound graph $SPG_k^u(s,t)$ is shown in Figure~\ref{fig:searchspace-edgelabel}(c).
Take edge $e(i,j)$ as an example. Since for $k_f=2$ and $k_b=3$ we have $EV^\ast_{k_f}(s,i)=\{ s,a,i \}$ and $EV^\ast_{k_b}(j,t)=\{j,h,b,t\}$ according to Figure~\ref{fig:EV}(a)-(b). $EV^\ast_{k_f}(s,i) \cap EV^\ast_{k_b}(j,t)=\emptyset$ leads to $e(i,j)\in SPG^u_k$.
In contrast, edge $e(b,j)$ is a failing edge since $EV^\ast_{k_b}(j,t)=\{j,h,b,t\}$ for $k_b \geq 3$, but $b \in EV^\ast_{k_f}(s,b)$ for $k_f \geq 2$ and $EV^\ast_{k_f}(s,b) \cap EV^\ast_{k_b}(j,t)\neq\emptyset$.
\end{example}

When examining each edge $e(u,v)$, it is unnecessary to enumerate all integer pairs ($k_f$, $k_b$) s.t. $k_f+1+k_b \leq k$.
Actually, we iterate $k_f$ from $0$ to $k-1$ and check requirements over $EV^\ast_{k_f}(s,u)$ and $EV^\ast_{k_b}(v,t)$ with $k_b=k-k_f-1$.
It is because checking $k_b=k-k_f-1$ already covers the examination for all $k_b<k-k_f-1$.
The correctness is guaranteed by Theorem~\ref{the: judgement cover}.

\begin{theorem}
\label{the: judgement cover}
For a certain $k_f$ {($0\leq k_f \leq k-1$)} and $k_b= k-k_f-1$, if $EV_{k_b}^\ast(v,t)$ does not exist,  $EV_{l}^\ast(v,t)$ will not exist for any $l< k_b$. And if $EV_{k_f}^\ast(s,u)\cap EV_{k_b}^\ast(v,t)\neq \emptyset$, $EV_{k_f}^\ast(s,u)\cap EV_{l}^\ast(v,t)\neq \emptyset$.
\end{theorem}

\begin{proof}
Since $P_k^\ast(v,t)$ denotes all $k$-hop-constrained $s$-$t$ simple paths, $\forall l<k_b$, $P_l^\ast(v,t)\subseteq P_{k_b}^\ast(v,t)$, and $EV_{k_b}^\ast(v, t) \subseteq EV_l^\ast(v, t)$.
If $EV_{k_b}^\ast(v,t)$ does not exist, $P_{k_b}^\ast(v,t)=\emptyset$.
Thus, $P_l^\ast(v,t)=\emptyset$, $EV_{l}^\ast(v,t)$ does not exist. 
When $EV_{k_b}^\ast(v,t)$ exists and $EV_{k_f}^\ast(s,u)\cap EV_{k_b}^\ast(v,t)\neq \emptyset$, as $EV_{k_b}^\ast(v, t)\subseteq EV_l^\ast(v, t)$, $EV_{k_f}^\ast(s,u)\cap EV_{l}^\ast(v,t)\neq \emptyset$.
\end{proof}

In practice, Algorithm~\ref{Edge Labeling} prunes failing edges by labeling them with ``0'', and forms the upper-bound graph $SPG_k^u(s,t)$ consisting of both undetermined edges (label ``1'') and definite edges (label ``2'').
Lines~\ref{line: 1-hop start}-\ref{line: 2-hop,t} work for $k_f \in \{0, 1, k-2, k-1\}$,
labeling ``2'' for definite edges which will be discussed in Section~\ref{sec: answer edge}.
Lines~\ref{line: tf}-\ref{line: tb binds with tf} work for $2\leq k_f \leq k-3$, 
when conditions in Definition~\ref{def:upper bound graph} are satisfied, it returns label ``1'' in lines~\ref{line: satisfy constraints}-\ref{line: label 1} (i.e., $e(u,v)$ is an undetermined edge).
Otherwise, label ``0'' is returned and $e(u,v)$ is a failing edge.

\subsection{Identifying Definite Edges}
\label{sec: answer edge}

After pruning failing edges, edges in $SPG_k^u(s,t)$ are further divided into \textit{definite edges} and \textit{undetermined edges}. 
Definite edges are definitely contained in the $SPG_k(s,t)$ by easy examination, while undetermined edges need further verification.
If more definite edges can be identified, the verification cost will
be reduced significantly. 

As shown in step (ii) of the overview (Figure~\ref{fig:overview}(b)), definite edges are those within two-hops from $s$ or to $t$ in $SPG_k^u(s,t)$.
Lemma~\ref{lemma:one hop} shows that the first-hop edges from $s$ (e.g., Figure~\ref{fig:searchspace-edgelabel}(d)) are definitely in $SPG_k(s,t)$, followed by Lemma~\ref{lemma:two hops} for those second-hop edges from $s$ (e.g., Figure~\ref{fig:searchspace-edgelabel}(e)).

\begin{algorithm} [t] 
	\caption{Edge Labeling}
	\footnotesize
	\label{Edge Labeling}
	\begin{algorithmic}[1]
		\Require Query variables ($s$, $t$ $k$), essential vertices $EV^\ast(s,u)$ and $EV^\ast(v,t)$, and edge $e(u,v)$.
		\Ensure Label of edge $e(u,v)$.
		\If{ ($u=s$ $\land$ ($EV_{k-1}^\ast(v,t)$ exists)) $\lor$ ($v=t$ $\land$ ($EV_{k-1}^\ast(s,u)$ exists))} \label{line: 1-hop start}
		    \State \Return 2 \label{line: 1-hop}
		\EndIf
		
		\If{($EV_{1}^\ast(s,u)$ and $EV_{k-2}^\ast(v,t)$ exist) $\land$ $u\notin EV_{k-2}^\ast(v,t)$}
		    \Return 2 \label{line: 2-hop,s}
		\EndIf
  
		\If{($EV_{1}^\ast(v,t)$ and $EV_{k-2}^\ast(s,u)$ exist) $\land$ $v\notin EV_{k-2}^\ast(s,u)$}
		    \Return 2 \label{line: 2-hop,t}
		\EndIf
		        
		\For{$k_f=2$ to $k-3$} \label{line: tf}
    		\State $k_b \leftarrow k-k_f-1$ \label{line: tb binds with tf}
    		\If{ ($EV_{k_f}^\ast(s,u)$ exists) $\land$ ($EV_{k_b}^\ast(v,t)$ exists)} \label{line: satisfy constraints}
        		\If{  $EV_{k_f}^\ast(s,u) \cap EV_{k_b}^\ast(v,t) =\emptyset$}
        		    \Return 1 \label{line: label 1}
        		\EndIf
    		\EndIf
		\EndFor
				
		\State \Return 0
	\end{algorithmic} 
\end{algorithm}

\begin{lemma}
\label{lemma:one hop}
$e(s,v)$ and $EV_{k-1}^\ast(v,t)$ exist $\Leftrightarrow$ $e(s,v) \in SPG_k(s,t)$.
\end{lemma}

\begin{proof}
If $EV_{k-1}^\ast(v,t)$ exists, $\exists \; p^\ast(v,t)$ with $s \notin V(p^\ast(v,t))$ and $\lvert p^\ast(v,t) \rvert \leq k-1$. 
Combining $p^\ast(v,t)$ with edge $e(s,v)$ constitutes a simple path $p^\ast(s,t)$ with $\lvert p^\ast(s,t) \rvert \leq k$. Thus, $e(s,v) \in SPG_k(s,t)$.

If $e(s,v) \in SPG_k(s,t)$, there exists a simple path $p^\ast(s,t)$ through edge $e(s,v)$ with $\lvert p^\ast(s,t) \rvert \leq k$. This also means $\exists $ $p^\ast(v,t)$ with $\lvert p^\ast(v,t) \rvert$  $\leq k-1$. Thus, $P_{k-1}^\ast(v,t)\neq \emptyset$ and $EV_{k-1}^\ast(v,t)$ exists. 
\end{proof}

\begin{example}
For graph $G$ in Figure~\ref{fig:searchspace-edgelabel}(c) and $k=7$, edge $e(s,a)\in SPG_k$ since $EV_{k-1}^\ast(a,t)$ exists according to Figure~\ref{fig:EV}(b).
\end{example}

\begin{lemma}
\label{lemma:two hops}
When both $EV^\ast_1(s,u)$ and $EV^\ast_{k-2}(v,t)$ exist, it holds that $EV^\ast_1(s,u) \cap EV^\ast_{k-2}(v,t)$ $=\emptyset$ 
$\Leftrightarrow$ $e(u,v)\in SPG_k(s,t)$.
\end{lemma}

\begin{proof}
When $EV^\ast_1(s,u)$ exists, $P_1^\ast(s,u)$ only contains a one-hop path $p^\ast=\{s,u\}$. 
Thus, $EV^\ast_1(s,u) = \{s,u\}$.  

If $EV^\ast_1(s,u) \cap EV^\ast_{k-2}(v,t)=\emptyset$, there must exist a simple path $p^\ast(v,t)$ without going through vertex $s$ and $u$, otherwise we have $u \in EV^\ast_{k-2}(v,t)$ since $s$ is excluded when computing $EV^\ast_{k-2}(v,t)$. 
Combining $p^\ast(v,t)$ with edge $e(u,v)$ and $e(s,u)$ constitutes a simple path $p^\ast(s,t)$ with $\lvert p^\ast(s,t) \rvert \leq k$. Thus, $e(u,v) \in SPG_k(s,t)$.

If $e(u,v) \in SPG_k(s,t)$, there must exist a simple path $p^\ast(s,t)$ through $e(u,v)$ with $\lvert p^\ast(s,t) \rvert \leq k$. 
Assume that $p^\ast(s,t)$ is composed of $p^\ast(s,u)$, $e(u,v)$ and $p^\ast(v,t)$.
Obviously, $u \notin V(p^\ast(v,t))$. As $\lvert p^\ast(v,t) \rvert \leq k-2$, $u \notin EV^\ast_{k-2}(v,t)$. 
Since $EV^\ast_1(s,u)=\{s,u\}$ and $s \not\in EV^\ast_{k-2}(v,t)$, $EV^*_1(s,u) \cap EV^\ast_{k-2}(v,t)=\emptyset$.
\end{proof}

\begin{example}
For graph $G$ in Figure~\ref{fig:searchspace-edgelabel}(c) and $k=7$, edge $e(a,i)\in SPG_k$ since $EV_{1}^\ast(s,a)=\{s,a\}$ and $EV_{k-2}^\ast(i,t)=\{i,j,h,b,t\}$ according to Figure~\ref{fig:EV}(a)-(b), indicating that $EV_{1}^\ast(s,a) \cap EV_{k-2}^\ast(i,t)=\emptyset$.
\end{example}
 
Based on Lemma \ref{lemma:one hop} and Lemma \ref{lemma:two hops}, it is clear that in upper-bound graph $SPG_k^u(s,t)$, all edges within two-hops from $s$ are definite edges. 
Similar conclusion can be reached that all edges within two-hops to $t$ are definite edges. 
Algorithm~\ref{Edge Labeling} therefore first tries to identify these definite edges. 
It checks whether (1) $u=s$ and $EV_{k-1}^\ast(v,t)$ exists, or (2) $v=t$ and $EV_{k-1}^\ast(s,u)$ exists. 
If one of the constraints is satisfied, it sets the indicating label of edge $e(u,v)$ to 2 (line~\ref{line: 1-hop}). 
Then, it checks whether $EV_1^\ast(s,u)$ and $EV_{k-2}^\ast(v,t)$ exist. 
If both of them exist and $u \notin EV_{k-2}^\ast(v,t)$ (i.e., $EV^*_1(s,u) \cap EV^\ast_{k-2}(v,t)=\emptyset$), it sets the label of $e(u,v)$ to $2$. 
Similar examination on $EV_1^\ast(v,t)$ follows (lines~\ref{line: 2-hop,s}-\ref{line: 2-hop,t}).  If none of them holds, it checks whether $e(u,v)$ is an undetermined edge by Definition~\ref{def:upper bound graph}.

\subsection{Further Analysis}
Based on definite edges, we then present two important properties as revealed in the following theorems. 

\begin{theorem}\label{cor:query4}
For all the queries with $k \leq 4$, the upper-bound graph is exactly the answer, i.e., $SPG^u_k(s,t)=SPG_k(s,t)$.
\end{theorem}

\begin{proof}
When $k \leq 4$, assume that there exists an edge $e(u,v)$ in $SPG^u_k(s,t)$ but not in $SPG_k(s,t)$.
Note that Algorithm~\ref{Edge Labeling} cannot assign label $1$ for $e(u,v)$ in line~\ref{line: tf}-\ref{line: label 1}.
If Algorithm~\ref{Edge Labeling} is returned in line~\ref{line: 1-hop}, then $e(u,v) \in SPG_k(s,t)$ by Lemma~\ref{lemma:one hop}.
Otherwise, Algorithm~\ref{Edge Labeling} is returned in lines~\ref{line: 2-hop,s}-\ref{line: 2-hop,t}, and $e(u,v) \in SPG_k(s,t)$ by Lemma~\ref{lemma:two hops}.
Since it contradicts the assumption, we have $SPG^u_k(s,t)=SPG_k(s,t)$.
\end{proof}

\begin{theorem}
\label{the:query4}
For any simple path $p^\ast(s,t)$ of length $l$, its first two edges and last two edges are definite edges.
\end{theorem}

\begin{proof}
Suppose edge $e(u,v)$ is the $i^{th}$ one in $p^\ast(s,t)$. 
If $i=1$ or $l$, we have $e(u,v) \in SPG_k(s,t)$ based on Lemma~\ref{lemma:one hop}. 
When $l>2$, if $i=2$ or $l-1$, then $e(u,v) \in SPG_k(s,t)$ based on Lemma~\ref{lemma:two hops}. 
Thus, $e(u,v)$ is a definite edge.
\end{proof}

When $k > 4$, it is not guaranteed that $SPG^u_k(s,t)=SPG_k(s,t)$, but all edges falling in the first and the last two steps of path $p^\ast(s,t)$ have been identified as definite edges by Theorem~\ref{the:query4}.
This important property significantly narrows the search space and will benefit the subsequent verification for the undetermined edges.

\textbf{Time complexity of Edge Labeling.} 
In Algorithm~\ref{Edge Labeling}, we examine each edge independently. 
Since the size of each essential vertex set is bounded by $k$, identifying definite edges in lines~\ref{line: 1-hop start}-\ref{line: 2-hop,t} takes $\mathcal{O}(k)$ time.
And by iterating $k_f$, it takes $\mathcal{O}(k^2)$ time for set intersection in lines~\ref{line: tf}-\ref{line: label 1}.
Thus, its total time complexity is $\mathcal{O}(k^2 \lvert E \rvert)$. 

\textbf{Space complexity of Edge Labeling.} 
Only the label of each edge will be maintained. 
Thus, the space complexity is $\mathcal{O}(\lvert E \rvert)$.

\section{Verifying Undetermined Edges}
\label{sec:Verification For Uncertain Edges} 

This section aims at the third step (of the overview, Figure~\ref{fig:overview}(c)) that obtains exact $SPG_k(s,t)$ from $SPG^u_k(s,t)$ by verifying undetermined edges.
We assume that hop constraint $k \geq 5$ in the following discussion, since $SPG_k(s,t)=SPG^u_k(s,t)$ if $k \leq 4$ (Theorem~\ref{cor:query4}). 

\begin{figure}[t]
  \centering
  \includegraphics[width=\linewidth]{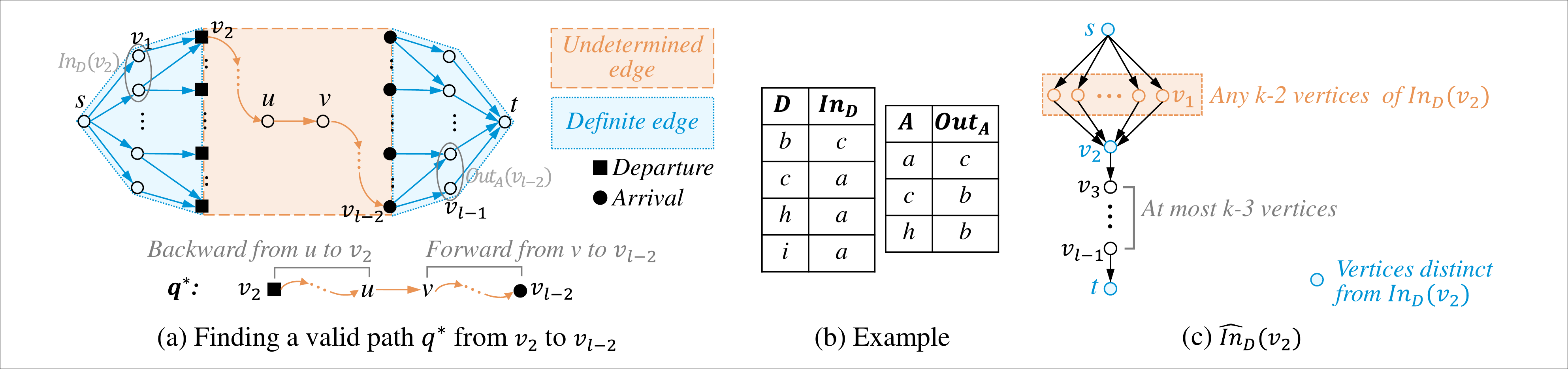}
  \caption{Illustrations for verifying undetermined edges}
  \label{fig:verification}
\end{figure}

Actually, finding one $k$-hop-constrained $s$-$t$ simple path through edge $e(u,v)$ is enough for concluding that $e(u,v) \in SPG_k(s,t)$.
As shown in Figure~\ref{fig:verification}(a), intuitively such a valid path passes through the boundary vertices between definite and undetermined edges.
Such boundary vertices are defined as \textit{Departure Vertex Set} $D$ (departures for short) and \textit{Arrival Vertex Set} $A$ (arrivals for short). A DFS-oriented algorithm with carefully designed search orders is developed to find valid paths from \textit{departure} to \textit{arrival}.

\subsection{Departures and Arrivals}
\label{sec:departures and arrivals}

Besides departures $D$ and arrivals $A$, we further introduce the concept of \textit{Valid In-neighbors of Departures $In_D$} (resp. \textit{Valid Out-neighbors of Arrivals $Out_A$}) that connect $s$ and  departure (resp. arrival and $t$).
As illustrated in Figure~\ref{fig:verification}(a), $v_2$ is a departure vertex and $v_{l-2}$ is an arrival vertex, and $v_1 \in In_D(v_2)$ ($v_{l-1} \in Out_A(v_{l-2})$) since it connects $s$ and $v_2$ ($v_{l-2}$ and $t$).

\begin{definition}
(Departure Vertex Set $D$).
\label{def: departures}
Vertex $v\in D$ iff there exists an in-neighbor $x$ of $v$ s.t. (1) $x$, $v$, $s$ and $t$ are distinct; and (2) edges $e(s,x)$ and $e(x,v)$ are contained in $SPG_k^u(s,t)$.
\end{definition}

\begin{definition}
\label{def: In_D}
(Valid In-neighbors of Departures, denoted as $In_D(v)$).
For each $v\in D$, vertex $x\in In_D(v)$ iff it makes the two conditions in Definition~\ref{def: departures} hold.
\end{definition}

\begin{definition}
(Arrival Vertex Set $A$).
\label{def: arrivals}
Vertex $v\in A$ iff there exists an out-neighbor $y$ of $v$ s.t. (1) $v$, $y$, $s$ and $t$ are distinct; (2) edges $e(v,y)$ and $e(y,t)$ are contained in $SPG_k^u(s,t)$.
\end{definition}

\begin{definition}
\label{def: Out_A}
(Valid Out-neighbors of Arrivals $Out_A(v)$).
For each $v\in A$, vertex $y\in Out_A(v)$ iff requirements (1) and (2) are satisfied in Definition~\ref{def: arrivals}.
\end{definition}

By Lemmas~\ref{lemma:one hop} and \ref{lemma:two hops}, edges $e(s,x)$, $e(x,v)$ in Definition~\ref{def: departures} and edges $e(v,y)$, $e(y,t)$ in Definition~\ref{def: arrivals} are definite edges.

\begin{example}
Consider the graph $G$ in Figure~\ref{fig:searchspace-edgelabel}(c), when $k=7$, its departure vertex set $D$ and arrival vertex set $A$ are shown in Figure~\ref{fig:verification}(b). 
For departure vertex $c$, its valid in-neighbor $In_D(c)=\{ a \}$ since edges $e(s,a)$, $e(a,c)\in SPG_k^u(s,t)$. 
Meanwhile, $c$ is also an arrival and its valid out-neighbor $Out_A(c)=\{b\}$.
\end{example}

Departure vertex set $D$ and their valid in-neighbors $In_D$ can be collected with second-hop edges (as illustrated in Figure~\ref{fig:searchspace-edgelabel}(e)). 
Notice that in  Algorithm~\ref{Edge Labeling}, when both $EV^\ast_1(s,u)$ and $EV_{k-1}^\ast(v,t)$ exist, and $u\notin EV_{k-2}^\ast(v,t)$, it labels edge $e(u,v)$ as a definite edge (line~\ref{line: 2-hop,s}). 
When these constraints are met, $s$, $u$, $v$, and $t$ are distinct and $e(s,u)$ is contained in $SPG_k^u(s,t)$, which indicates that $v$ is a departure vertex and $u$ is a valid in-neighbor of $v$. 
Hence, to obtain departures $D$ and $In_D$, we only need to add such $v$ into $D$ and $u$ into $In_D(v)$ while labeling edges in line~\ref{line: 2-hop,s}.
Similarly, arrivals $A$ and $Out_A$ can also be collected.

\begin{algorithm}[t]
	\caption{Undetermined Edge Verification}
	\label{alg: Uncertain Edge Validation}
	\footnotesize
	\begin{algorithmic}[1]
		\Require Query variables ($s$, $t$, $k$), upper-bound graph $SPG_k^u$, departures $D$, arrivals $A$, valid in-neighbors $In_D$, and valid out-neighbors $Out_A$.
		\Ensure All edges in $SPG_k$.
		\State $E(SPG_k) \leftarrow$ All definite edges in $SPG_k^u$ \label{line: initialize with answer edges}
		\If{$k \geq 5$} \label{line: start validation}
    		\For {each undetermined edge $e(u,v) \not\in E(SPG_k)$}
    		    \State $stk_v \leftarrow \{u,v,s,t\}$, $stk_e \leftarrow \{e(u,v)\}$
    		    \State Forward($v$, $1$, $u$) \label{line: invoke forward function}
    		\EndFor
		\EndIf \label{line: end validation}
		\State \Return $E(SPG_k)$ \label{line: return SPG_k}
	    \Function{Forward}{$cur$, $l$, $u$}
    	    \If{$cur \in A$ $\land$ Backward($u$, $l$, $cur$)} \label{line: invoke backward function}
    		     \Return $True$
    		\EndIf
    		\If{$l<k-4$} \label{line: forward length bound}
        		\For{each out-edge $e(cur,nxt)$ in $SPG_k^u$ s.t. $nxt \not\in stk_v$} \label{line: start forward exploration}
        		    \State Push $nxt$ into $stk_v$, push $e(cur,nxt)$ into $stk_e$
        		    \If{Forward($nxt$, $l+1$, $u$)}
        		        \Return $True$
        		    \EndIf
        		    \State Pop $nxt$ from $stk_v$, pop $e(cur,nxt)$ from $stk_e$
        		\EndFor
        	\EndIf \label{line: end forward exploration}
        	\State \Return $False$
		\EndFunction
		\Function{Backward}{$cur$, $l$, $arrival$}
    	    \If{$cur \in D$ $\land$ TryAddEdges($cur$, $arrival$)} \label{line: invoke TryAddEdges}
    		    \Return $True$
    		\EndIf
    		\If{$l<k-4$} \label{line: backrward length bound}
        		\For{each in-edge $e(nxt,cur)$ in $SPG_k^u$ s.t. $nxt \not\in stk_v$} \label{line: start backward exploration}
        		    \State Push $nxt$ into $stk_v$, push $e(nxt,cur)$ into $stk_e$
        		    \If{Backward($nxt$, $l+1$, $arrival$)}
        		        \Return $True$
        		    \EndIf
        		    \State Pop $nxt$ from $stk_v$, pop $e(nxt,cur)$ from $stk_e$
        		\EndFor
        	\EndIf \label{line: end backward exploration}
        	\State \Return $False$
		\EndFunction
		\Function{TryAddEdges}{$departure$, $arrival$}
            \State $In_D^c \leftarrow \{x \mid x \in In_D(departure) \land x \not\in stk_v\}$ \label{line: In_c}
            \State $Out_A^c \leftarrow \{y \mid y \in Out_A(arrival) \land y \not\in stk_v\}$ \label{line: Out_c}
    		\If{$\exists x \in In_D^c, \; \exists y \in Out_A^c$ s.t. $x \neq y$} \label{line: have distinct vertex}
    		    \State $E(SPG_k) \leftarrow E(SPG_k) \cup stk_e$, \Return $True$
    		\EndIf
    		\State \Return $False$
    	\EndFunction
	\end{algorithmic} 
\end{algorithm}

\subsection{DFS-oriented Search}

As shown in Figure~\ref{fig:verification}(a), for concluding that the undetermined edge $e(u,v)\in SPG_k$, intuitively we need to find a simple path $q^\ast$ from departure $v_2$ to arrival $v_{l-2}$ through edge $e(u,v)$, s.t. joining $s$, $v_1$, $q^\ast$, $v_{l-1}$, and $t$ produces a desired $s$-$t$ simple path. 
We then formalize the requirements for $q^\ast$ into the following theorem.

\begin{theorem}
\label{the: verification requirements}
Undetermined edge $e(u,v) \in SPG_k(s,t) \Leftrightarrow$ there exists a simple path $q^\ast=\{v_2, v_3, \ldots,$ $v_{l-2}\}$ through $e(u,v)$ within $k-4$ hops, where (1) $v_2 \in D$ $\land$ $v_{l-2} \in A$; and (2) $\exists v_1 \in In_D(v_2)$, $\exists v_{l-1} \in Out_A(v_{l-2})$ s.t. vertices $s, v_1, v_2, \ldots, $ $v_{l-1}, t$ are distinct.
\end{theorem}

\begin{proof}
Undetermined edge $e(u,v)$ is contained in $SPG_k(s,t)$ iff there exists an $s$-$t$ simple path $p^\ast=\{s=v_0, v_1, \ldots, v_{l-1}, v_l=t\}$ through $e(u,v)$ s.t. $|p^\ast|=l \leq k$.
Note that $e(u,v)$ is not in the first or last two steps of $p^\ast$, or it will not be an undetermined edge.

\noindent \textbf{Necessity}.
Joining $e(s,v_1)$, $e(v_1, v_2)$, edges in $q^\ast$, $e(v_{l-2}, v_{l-1})$ and $e(v_{l-1},t)$ in order forms such a simple path $p^\ast$.

\noindent \textbf{Sufficiency}. 
$q^\ast$ can be extracted from $p^\ast$, which starts from $v_2$ and ends at $v_{l-2}$ with length $l-4 \leq k-4$.
By definitions, $v_2 \in D$, $v_{l-2} \in A$, $v_1 \in In_D(v_2)$ and $v_{1-1} \in Out_A(v_{l-2})$.
And vertices $s, v_1, v_2, \ldots, v_{l-1}, t$ are distinct since they are in simple path $p^\ast$.
\end{proof}

Based on Theorem~\ref{the: verification requirements}, we present Algorithm~\ref{alg: Uncertain Edge Validation} to efficiently conclude whether an undetermined edge $e(u,v)$ is contained in $SPG_k(s,t)$.
First, $E(SPG_k)$ is initialized with all definite edges in line~\ref{line: initialize with answer edges}.
In lines~\ref{line: start validation}-\ref{line: end validation}, for each undetermined edge $e(u,v)$, DFS-oriented search is conducted to find a valid path $q^\ast$ described in Theorem~\ref{the: verification requirements}.
As illustrated in Figure~\ref{fig:verification}(a), it first tries to find a path from $v$ to an arrival $v_{l-2}$ (function \textit{Forward}) and a path from a departure $v_2$ to $u$ (function \textit{Backward}). The two paths are joined with $e(u,v)$, forming a path $q^\ast$ from departure $v_2$ to arrival $v_{l-2}$.
If $q^\ast$ passes verification in function \textit{TryAddEdges}, edges in $q^\ast$ will be added into $E(SPG_k)$.
Let us consider the following example.

\begin{example}
For graph $G$ in Figure~\ref{fig:searchspace-edgelabel}(c) with $k=7$, to verify the undetermined edge $e(i,j)$, we search forward from vertex $j$ and visit edge $e(j,h)\in SPG_k^u$.
Since an arrival vertex $h$ (see Figure~\ref{fig:verification}(b)) is reached, we search backward from $i$ and find that $i$ is a departure vertex.
DFS-oriented search terminates with a simple path $q^\ast =\{i,j,h\}$.
According to Figure~\ref{fig:verification}(b), since $In_D(i)=\{a\}$ and $Out_A(h)=\{b\}$ have distinct vertices from $q^\ast$ and $|q^\ast|\leq k-4$, all undetermined edges involved ($e(i,j)$ and $e(j,h)$) will be added to result set.
\end{example}

By Theorem~\ref{the: size of In_D}, for any departure vertex $v \in D$ we only need to store at most $k-2$ vertices in $In_D(v)$, i.e., replacing $In_D(v)$ with $\widehat{In}_D(v)$, where $|In_D(v)|$ denotes the size of $In_D(v)$ and
\begin{equation*}
    \widehat{In}_D(v) = \left\{ 
        \begin{array}{ll}
            \text{any $k-2$ vertices in $In_D(v)$}, & |In_D(v)| > k-2; \\
            In_D(v),                                &  \text{otherwise}.
        \end{array}\right. 
\end{equation*}

\begin{theorem}
\label{the: size of In_D}
Theorem~\ref{the: verification requirements} holds if replacing $In_D$ with $\widehat{In}_D$.
\end{theorem}

\begin{proof}
We just need to consider the case that $|In_D(v_2)|>k-2$.
Recall that $In_D(v)$ is involved only in requirement (2) of Theorem~\ref{the: verification requirements}, which is $\exists v_1 \in In_D(v_2)$, $\exists v_{1-1} \in Out_A(v_{l-2})$ s.t. vertices $s, t, v_1, v_2, \ldots, v_{l-1}$ are distinct ($l \leq k$).
Without loss of generality, assume that vertices $s, t, v_2, v_3, \ldots, v_{l-1}$ are fixed and already distinct, as illustrated in Figure~\ref{fig:verification}(c).
We need to prove that given $|In_D(v_2)|>k-2$, $\exists v_1 \in In_D(v_2)$ s.t. $v_1$ is distinct from $s, t, v_2, v_3, \ldots, v_{l-1}$ iff $\exists v_1 \in \widehat{In}_D(v_2)$ s.t. $v_1$ is distinct from $s, t, v_2, v_3, \ldots, v_{l-1}$.

By Definition~\ref{def: In_D}, $\forall x \in \widehat{In}_D(v_2) \subseteq In_D(v_2)$, $x$ is distinct from $s$, $t$ and $v_2$.
To prove sufficiency, notice that number of the other vertices $v_3, v_4, \ldots, v_{l-1}$ is $l-3 \leq k-3$, and we have $k-2$ vertices in $\widehat{In}_D(v_2)$.
Hence, $\exists v_1 \in \widehat{In}_D(v_2)$ s.t. $v_1$ is distinct from $s, t, v_2, v_3, \ldots, v_{l-1}$.
Necessity is trivial since $\widehat{In}_D(v_2) \subseteq In_D(v_2)$.
\end{proof}

Similarly, at most $k-2$ vertices of $Out_A(v)$ will be materialized for any arrival vertex $v \in A$.

\textbf{Time complexity of Verification}.
Recall that $D$, $A$, $In_D$ and $Out_A$ can be collected within $\mathcal{O}(|E|)$ time.
For each undetermined edge $e(u,v)$, DFS-oriented search tries to find a simple path $q^\ast$ through $e(u,v)$ within $k-4$ hops, which takes $\mathcal{O}(d_{max}^{k-5})$ time and $d_{max}$ is the largest in-degree or out-degree of vertices in $SPG_k^u$.
Further verification in function \textit{TryAddEdges} for each $q^\ast$ takes $\mathcal{O}(k)$, since at most $k-2$ vertices are stored in $\widehat{In}_D(v_2)$ ($\widehat{Out}_A(v_{l-2})$) after replacing $In_D(v_2)$ ($Out_A(v_{l-2})$).
Therefore, verifying undetermined edges takes $\mathcal{O}(k|E|d_{max}^{k-5})$ in the worst.

\textbf{Space complexity of Verification}.
The number of departures and arrivals are bounded by $\mathcal{O}(|V|)$, and for each departure (arrival) $v$, $|\widehat{In}_D(v)| \leq k-2$ ($\widehat{Out}_A(v) \leq k-2$). 
Stacks for vertices and edges have size $\mathcal{O}(k)$.
Hence, the total space cost is $\mathcal{O}(k|V|)$.

\begin{theorem}
\label{the:time complexity k<=5}
For $k \leq 5$, \textit{EVE} takes $\mathcal{O}(|E|)$ time and space.
\end{theorem}

\begin{proof}
As discussed in Section~\ref{sec: Essential Vertex}, computing essential vertices takes $\mathcal{O}(k^2|E|)$ time and $\mathcal{O}(k^2|V|)$ space. 
In Section~\ref{sec: upperbound}, labeling edges and generating upper-bound graph consume $\mathcal{O}(k^2|E|)$ time and $\mathcal{O}(|E|)$ space, while edge verification takes $\mathcal{O}(k|E|)$ time and $\mathcal{O}(k|V|)$ space when $k=5$.
Therefore, both time and space complexity are $\mathcal{O}(|E|)$ by considering $k$ as a small constant when $k \leq 5$.
\end{proof}

\subsection{Search Ordering Strategies}
\label{sec: search ordering strategies}

As shown in Section~\ref{sec:Coverage Ratio and Redundant Ratio}, $SPG^u_k$ is a tight upper-bound of the desired simple path graph $SPG_k$, i.e., for each undetermined edge, a valid simple path $q^\ast$ discussed in Theorem~\ref{the: verification requirements} probably exists.
To further speed up DFS-oriented search, it is better to find a valid $q^\ast$ earlier, stop the search and return $True$ directly.
Note that when $k=5$ such techniques are unnecessary, since the initial length $l=k-4=1$ indicates that neither forward nor backward exploration is needed.

Recall requirements in Theorem~\ref{the: verification requirements}. To verify edge $e(u,v)$, we need vertices $v_2 \in D$, $v_{l-2} \in A$, $v_1 \in In_D(v_2)$ and $v_{l-1} \in Out_A(v_{l-2})$, while vertex distinctness and length constraint of $q^\ast$ are considered too.
Intuitively, when starting a forward search from $v$, if vertices closer to any arrival are explored first, the desired path $q^\ast$ within length constraint is more likely to be obtained.
Moreover, when there are several arrivals in the next step, visiting the one with larger $|Out_A(v_{l-2})|$ helps to increase the chance of satisfying vertex distinctness.
Such intuitions drive us to sort out-neighbors of each vertex in $SPG_k^u$ before conducting a DFS-oriented search.
We sort them in ascending order of distance to the closest arrival vertex, and those with distance $0$ (i.e., arrivals) are sorted by size of $Out_A$ in descending order.
Similarly, we can sort in-neighbors of each vertex in $SPG_k^u$ in ascending order of distance from the closest departure vertex, and those with distance $0$ (i.e., departures) are sorted by size of $In_D$ in descending order.

Calculating the distance from departures to vertices $w$ in $SPG_k^u$ costs $\mathcal{O}(|E|)$ time and space, since we can create a virtual vertex $r$ connecting to all departures, after which BFS is conducted from $r$ for obtaining such distances.
Sorting in-neighbors of all vertices takes $\mathcal{O}( \sum_{w \in V(SPG_k^u)} d_w \log (d_w)) = \mathcal{O}(|E| \log (d_{max}))$, where $d_w$ denotes the in-degree of vertex $w$ in $SPG_k^u$.
Since the same costs also hold for sorting out-neighbors, pre-computation for adjusting search orders totally takes $\mathcal{O}(|E| \log (d_{max}))$ time and $\mathcal{O}(|E|)$ space.

\section{Experiments}
\label{sec:Experiments}

\subsection{Experimental Setup}

\textbf{Datasets.}
As summarised in Table~\ref{tab:datasets}, we use 15 real networks from a variety of domains, such as social networks, web graphs and biological networks, in the experiments. 
These graphs are downloaded from NetworkRepository\footnote{\url{https://networkrepository.com/networks.php}} \cite{NetworkRepository}, SNAP\footnote{\url{http://snap.stanford.edu/data/}} \cite{snapnets} and Konect\footnote{\url{http://konect.cc/networks/}}~\cite{konect}.
The number of vertices and edges ranges from thousands to billions.

\begin{table}[t]
\caption{Networks for Experiments}
\footnotesize
\begin{tabular}{cccccc}
\hline
\textbf{Name} & \textbf{Dataset}      & $\mathbf{|V|}$ & $\mathbf{|E|}$ & $\mathbf{d_{avg}}$ & \textbf{Type} \\ \hline
\textit{ps}   & econ-psmigr3          & 3.1K           & 540K           & 172                & Economic      \\
\textit{ye}   & bio-grid-yeast        & 6K             & 314K           & 52                 & Biological    \\
\textit{wn}   & bio-WormNet-v3        & 16K            & 763K           & 47                 & Biological    \\
\textit{uk}   & web-uk-2005           & 130K           & 12M            & 91                 & Web           \\
\textit{sf}   & web-Stanford          & 282K           & 13M            & 46                 & Web           \\
\textit{bk}   & web-baidu-baike       & 416K           & 3.3M           & 8                  & Web           \\
\textit{tw}   & twitter-social        & 465K           & 835K           & 2                  & Miscellaneous \\
\textit{bs}   & web-BerkStan          & 685K           & 7.6M           & 11                 & Web           \\
\textit{gg}   & web-Google            & 876K           & 5.1M           & 6                  & Web           \\
\textit{hm}   & bn-human-Jung2015     & 976K           & 146M           & 150                & Biological    \\
\textit{wt}   & wikiTalk              & 2.4M           & 5M             & 2                  & Miscellaneous \\
\textit{lj}   & soc-LiveJournal1      & 4.8M           & 68M            & 14                 & Social        \\
\textit{dl}   & dbpedia-link          & 18M            & 137M           & 7                  & Miscellaneous \\
\textit{fr}   & soc-friendster        & 66M            & 1.8B           & 28                 & Social        \\
\textit{hg}   & web-cc12-hostgraph    & 89M            & 2B             & 23                 & Web           \\ \hline
\end{tabular}
\label{tab:datasets}
\end{table}

\begin{figure}[t]
  \centering
  \includegraphics[width=\linewidth]{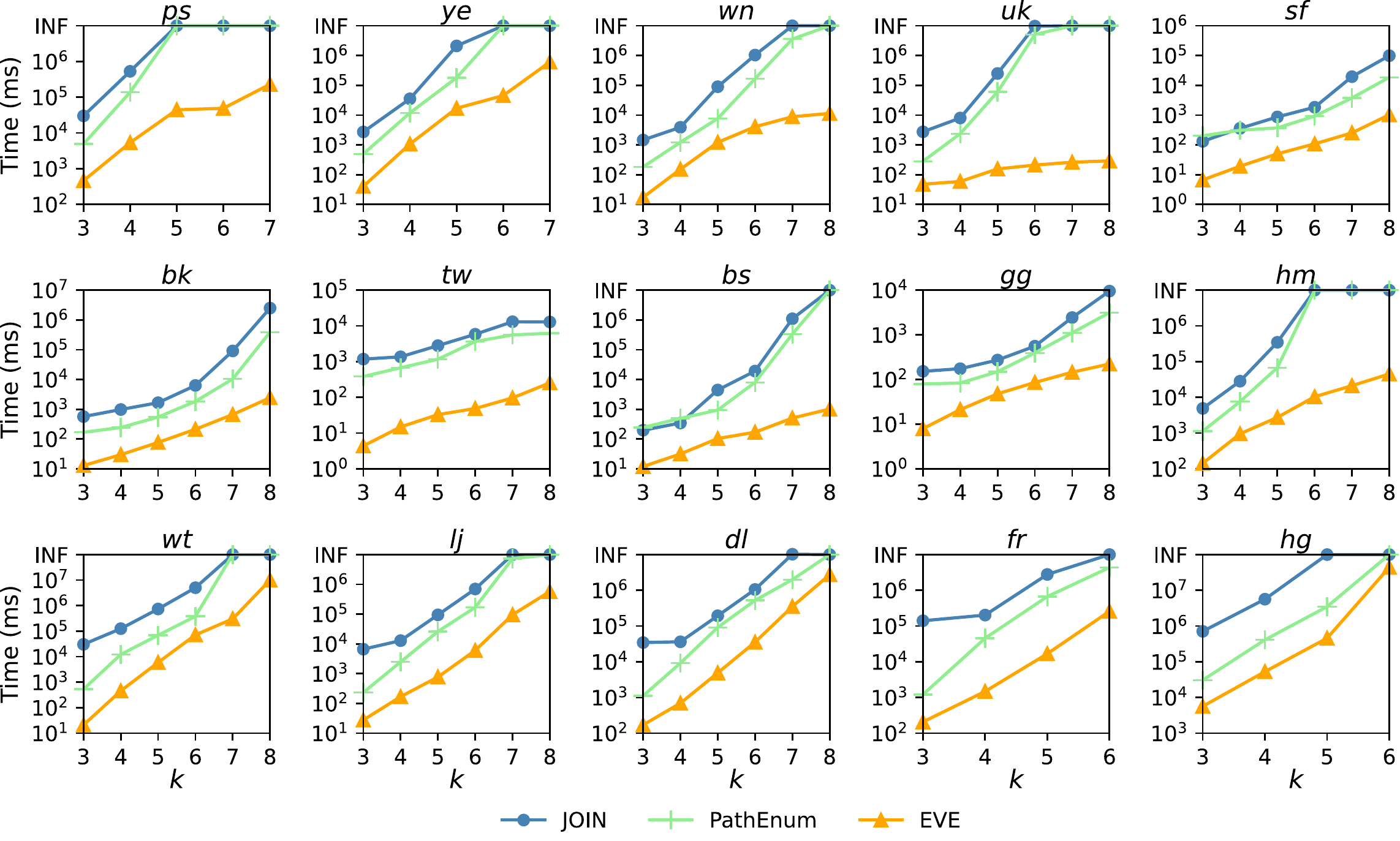}
  \caption{Total time cost of \textit{EVE} and baselines}
  \label{fig:TimeCostWithBaselines}
\end{figure}

\textbf{Queries.}
For each hop constraint $k$, we generate $1000$ random query pairs $(s,t)$ on each graph such that source vertex $s$ could reach target vertex $t$ in $k$ hops.
Note that we focus on $k$-hop reachable query pairs, since the others can be efficiently filtered out by answering $k$-hop reachability queries \cite{kReach2012, kReach2014, ESTI}.
Similar to previous works of path enumeration \cite{JOIN2020, JOIN2021, PathEnum}, large $k$ is also not considered, since relation strength usually drops dramatically w.r.t $k$ and long paths are not useful to capturing the tie between vertices \cite{JOIN2021}.

We conduct experiments on a Linux server with Intel(R) E5-2678v3 CPU @2.5GHz and 220G RAM.
Programs are implemented in C++ and compiled with -O3 Optimization. 

\subsection{Performance Comparison with Baselines}
\label{sec: compare baselines}

To generate $k$-hop-constrained $s$-$t$ simple path graph $SPG_k(s,t)$, enumerating all hop-constrained $s$-$t$ simple paths then put all edges and vertices of these paths together is a straightforward solution.
\textit{JOIN} \cite{JOIN2020, JOIN2021} and \textit{PathEnum} \cite{PathEnum} are state-of-the-arts for path enumeration, which significantly outperform other approaches and are thus considered as baseline algorithms for generating $SPG_k(s,t)$.
Note that given vertices $s$, $t$ and hop constraint $k$, the edges of each path found by \textit{JOIN} or \textit{PathEnum} will be inserted into a set $E(SPG_k(s,t))$ which will be returned as the answer.
Our proposed algorithm is denoted as \textit{EVE}.

\textbf{Results on time efficiency}.
Total time cost of answering $1000$ queries for each graph with different $k$ are shown in Figure~\ref{fig:TimeCostWithBaselines}, where the running time is set to \textit{INF} if an algorithm does not terminate in $1\times 10^7 $ ms ($1\times 10^8 $ ms for graphs \textit{wt} and \textit{hg}).
\textit{EVE} is clear the most efficient on all graphs and $k$, and generally at least an order of magnitude faster than baselines.
As expected, the time cost increases as $k$ grows for all algorithms. 
Moreover, the gap between \textit{EVE} and the baselines gets larger on dense graphs, such as \textit{ps}, \textit{ye}, \textit{wn}, \textit{uk} and \textit{hm}.
It is because paths' amount grows exponentially with large base of exponent, but edges' amount is bounded by $|E|$, as illustrated in Figure~\ref{fig:relfinder-pathcount}(b).
\textit{EVE} even outperforms baselines by $4$ orders of magnitude on graph \textit{uk} when $k \geq 6$.
The scalability of \textit{EVE} is also confirmed on billion-scale graphs \textit{fr} and \textit{hg}, while baselines may run out of time when $k=6$. For example, \textit{EVE} only takes $0.26$s to answer each query on average over graph \textit{fr} when $k=6$.

\begin{figure}[t]
  \centering
  \includegraphics[width=\linewidth]{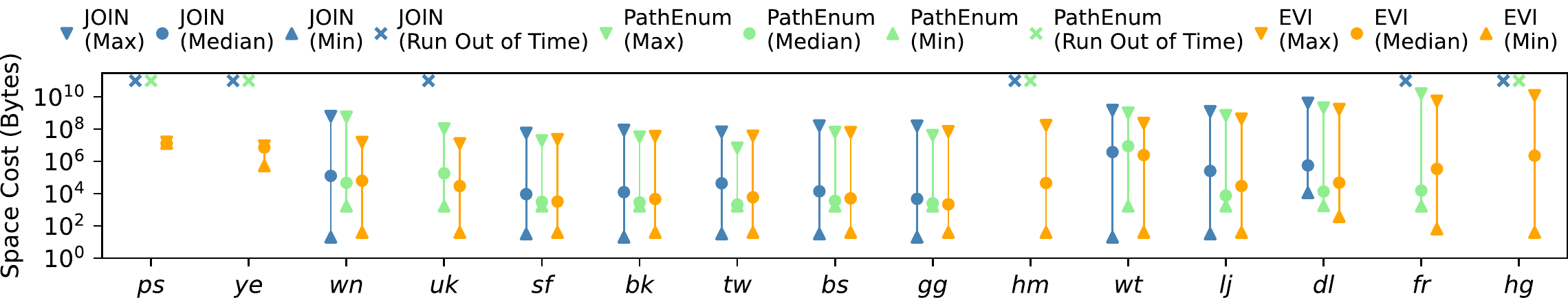}
  \caption{ Maximum, median, and minimum space cost among 1000 queries ($k=6$)}
  \label{fig:SpaceWithBaselines}
\end{figure}

\begin{figure}[t]
	\centering
	\subfloat[Maximum space cost for graphs \textit{wn} and \textit{bs}]{\includegraphics[width=.32\columnwidth]{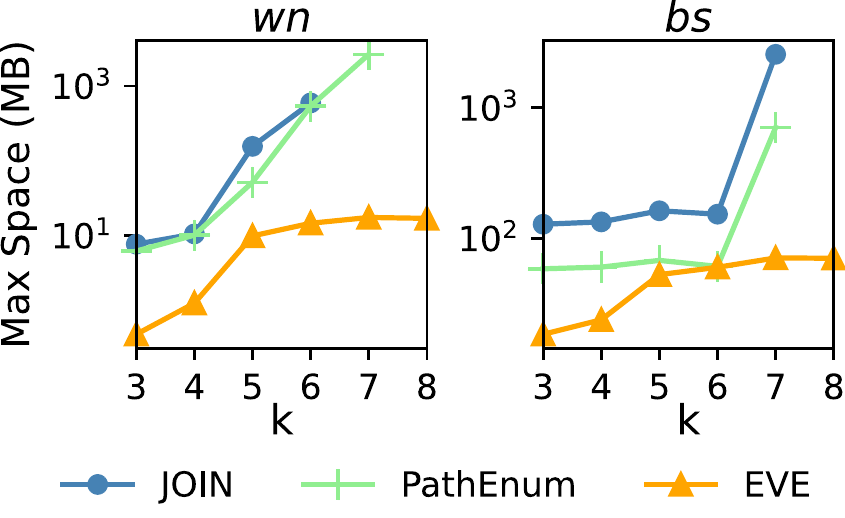}}\hspace{5pt}
	\subfloat[Average query time for varying distances between query vertices $s$ and $t$ on graphs \textit{lj} and \textit{bs} ($k=6$)]{\includegraphics[width=.32\columnwidth]{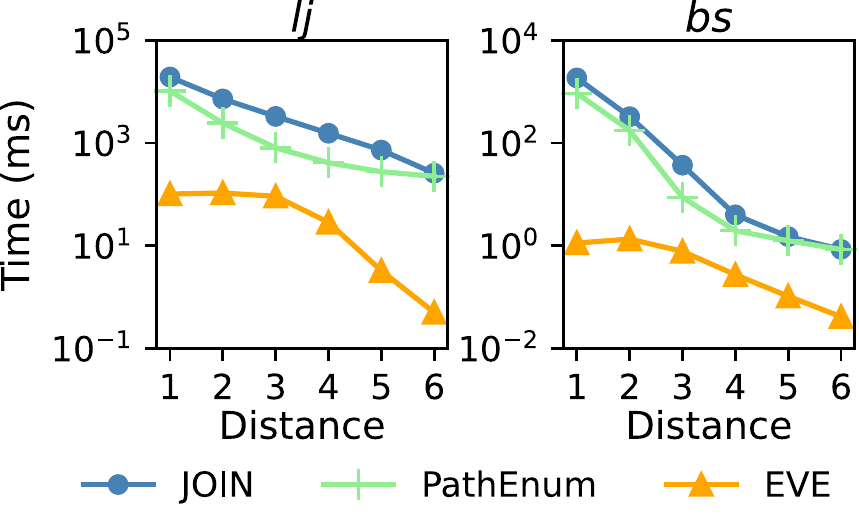}}\hspace{5pt}
    \subfloat[Detailed time cost for graphs \textit{ye} and \textit{bs}]{\includegraphics[width=.32\columnwidth]{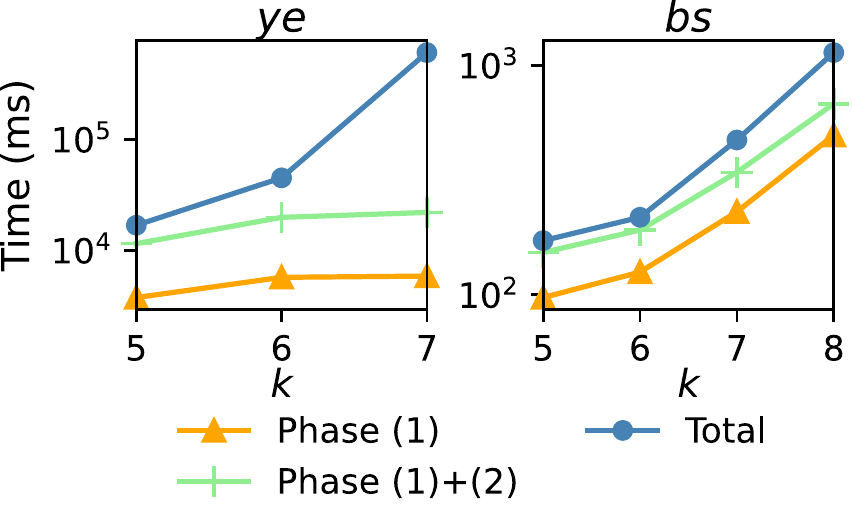}}
	\caption{Maximum space cost, average query time for varying distances, and detailed time cost}
    \label{fig:detail-analysis}
\end{figure}

\textbf{Results on space cost}.
Since space consumption varies for different queries, we report the maximum, median, and minimum space cost among $1000$ queries for each graph when $k=6$ in Figure~\ref{fig:SpaceWithBaselines}.
We then focus on maximum space cost since an algorithm needs large enough memory for certain queries.
\textit{JOIN} demands the largest space cost since partial paths are stored before joining to obtain an $s$-$t$ path.
\textit{PathEnum} takes less space since its pre-built index helps to reduce the number of partial paths when the join-based method is adopted. 
On graphs \textit{sf}, \textit{bk}, \textit{tw} and \textit{gg}, \textit{EVE} requires larger space compared with \textit{PathEnum}.
It is because these graphs have fewer desired paths, indicating that less space is needed to store partial paths for \textit{PathEnum}, while $\mathcal{O}(k^2|V|)$ space is consumed to store essential vertices for \textit{EVE}.
We further investigate how $k$ influences the maximum space cost by taking graphs \textit{wn} and \textit{bs} as examples, as shown in Figure~\ref{fig:detail-analysis}(a).
As $k$ increases, baselines take much larger space, since the number of paths grows exponentially and storing partial paths is space inefficient.
For \textit{EVE}, we notice that space cost grows rapidly from $k=4$ to $k=5$, since undetermined edges, departures, arrivals and their neighbors are maintained for verification when $k \geq 5$.

\subsection{Effect of Distances Between Query Pairs}
For different shortest distances $\Delta(s,t)$ of query pairs ($s,t$), \textit{EVE} is robust and significantly outperforms baselines.
Take graphs $lj$ and $bs$ as examples, when $k=6$ we generate 500 random queries for each distance $1, 2, \cdots, 6$.
As reported in Figure~\ref{fig:detail-analysis}(b), when $\Delta(s,t)$ is closer to $k$, the number of paths is likely to decrease, and less time is required for all algorithms.
\textit{EVE} always has much shorter running time, especially for small distances.
Intuitively, if $s$ and $t$ are closely connected, they tend to have more paths within $k$ hops.

\subsection{Detailed Time Cost of \textit{EVE} method}

We further investigate the detailed time cost of three phases in \textit{EVE}, i.e., (1) propagation for essential vertices, (2) computing upper-bound graph and (3) verifying undetermined edges.
Taking dense graph \textit{ye} and sparse graph \textit{bs} as examples, their detailed time cost for $k \geq 5$ is shown in Figure~\ref{fig:detail-analysis}(c).
As $k$ increases, the time cost for phase (3) grows rapidly in graph \textit{ye} due to its dense structure. 
For graph \textit{bs}, the first two phases dominate the total cost, i.e., the cost of verifying undetermined edges is marginal compared with the cost of generating the upper-bound graph.

\begin{table}[t]
\linespread{0.9} 
\footnotesize
\caption{Average redundant ratio $r_D$}
\label{tab:UPGdifferenceRatio}
\begin{tabular}{c|ccccccccccccccc}
\hline
\textbf{$k$} & \textit{ps} & \textit{ye} & \textit{wn} & \textit{uk} & \textit{st} & \textit{bk} & \textit{tw} & \textit{bs}  \\ \hline
        $5$ & 0.000004\% & 0.02\% & $0$ & $0$ & 0.49\% & 0.01\% & $0$ & 0.31\%  \\ 
        $6$ & $0$ & 0.009\% & $0$ & $0$ & 0.84\% & 0.03\% & $0$ & 0.45\%  \\ 
        $7$ & $0$ & 0.002\% & $0$ & $0$ & 1.4\% & 0.001\% & 0.004\% & 0.74\%  \\ 
        $8$ & - & - & $0$ & $0$ & 1.6\% & 0.02\% & 0.01\% & 0.95\%  \\ \hline
        \textbf{$k$} & \textit{gg} & \textit{hm} & \textit{wt} & \textit{lj} & \textit{dl} & \textit{fr} & \textit{hg} &   \\ \hline 
        $5$ & 0.09\% & $0$ & 0.00006\% & $0$ & $0$ & $0$ & $0$ &   \\ 
        $6$ & 0.12\% & $0$ & 0.001\% & 0.0002\% & 0.0003\% & $0$ & 0.002\% &   \\ 
        $7$ & 0.19\% & $0$ & 0.002\% & 0.001\% & 0.0009\% & - & - &   \\ 
        $8$ & 0.13\% & $0$ & 0.002\% & 0.004\% & 0.003\% & - & - &   \\ \hline
\end{tabular}
\end{table}

\begin{figure}[t]
  \centering
  \includegraphics[width=\linewidth]{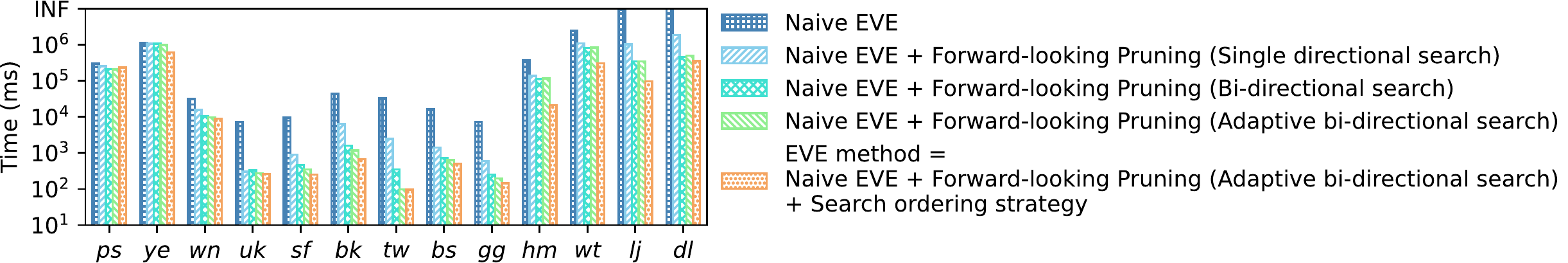}
  \caption{Effectiveness of pruning strategies for \textit{EVE} ($k=7$)}
  \label{fig:TimeCostWithPrunings}
\end{figure}

\subsection{Effectiveness of Pruning Strategies}
We evaluate the effectiveness of our proposed pruning techniques including {forward-looking pruning strategy} (Section~\ref{sec: adaptive bi-directional BFS}) and search ordering strategy (Section~\ref{sec: search ordering strategies}).
Figure~\ref{fig:TimeCostWithPrunings} reports the time cost of different versions that are equipped with distinct pruning techniques, where $k$ is set to $7$ by default and \textit{Naive EVE} is the one disabling all pruning techniques above.
With forward-looking pruning, running time on all graphs can be reduced up to an order of magnitude. The adaptive bi-directional search is faster than both single and bi-directional search.
The search ordering strategy  works effectively on most graphs except \textit{ps}, since the re-ordering cost is not paid off.
Note that \textit{ps} has the largest average degree $172$, and finding a valid path is much easier in such a dense graph.

\subsection{Coverage Ratio and Redundant Ratio}
\label{sec:Coverage Ratio and Redundant Ratio}

\begin{figure}[t]
	\centering
	\subfloat[Average coverage ratio $r_C$ vs. $k$]{\includegraphics[width=.515\columnwidth]{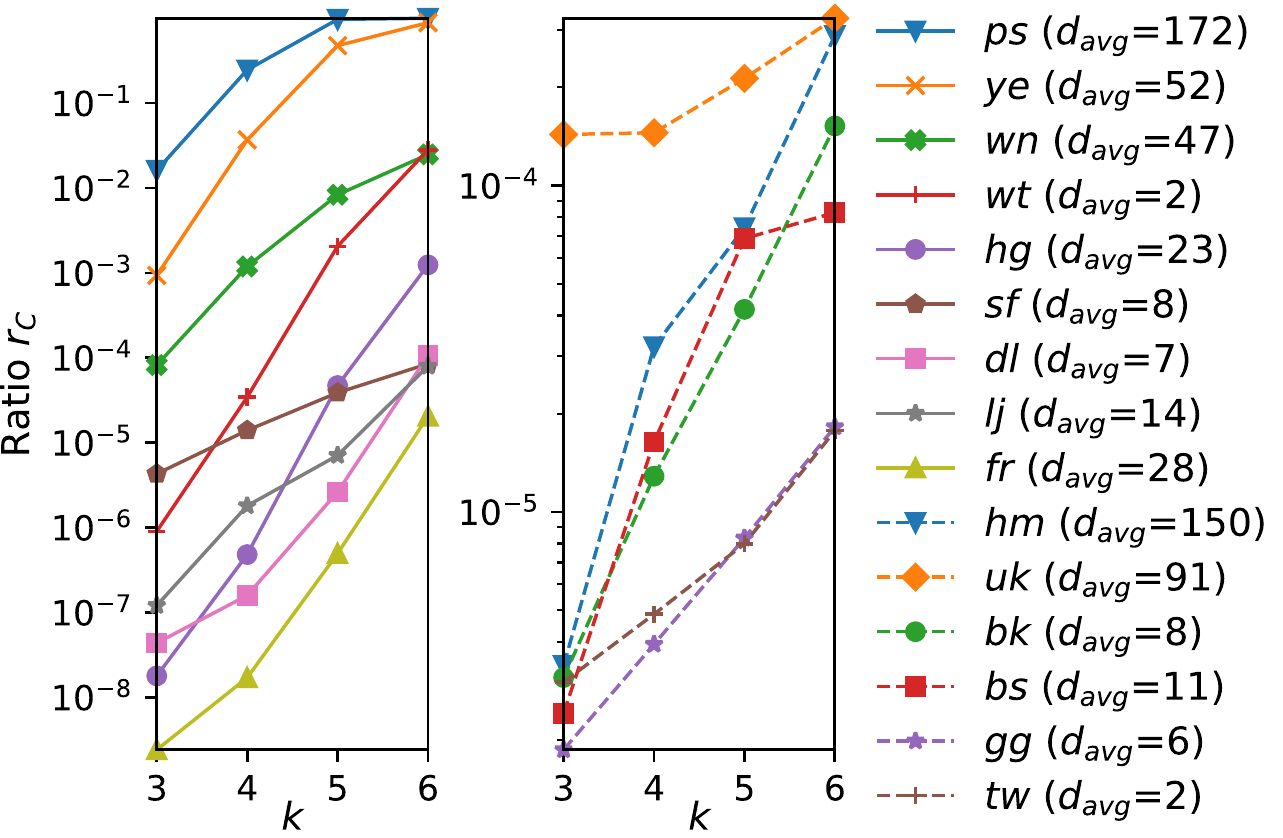}}\hspace{5pt}
	\subfloat[Compare with baselines on graphs \textit{tw}, \textit{lj}, and \textit{dl}]{\includegraphics[width=.465\columnwidth]{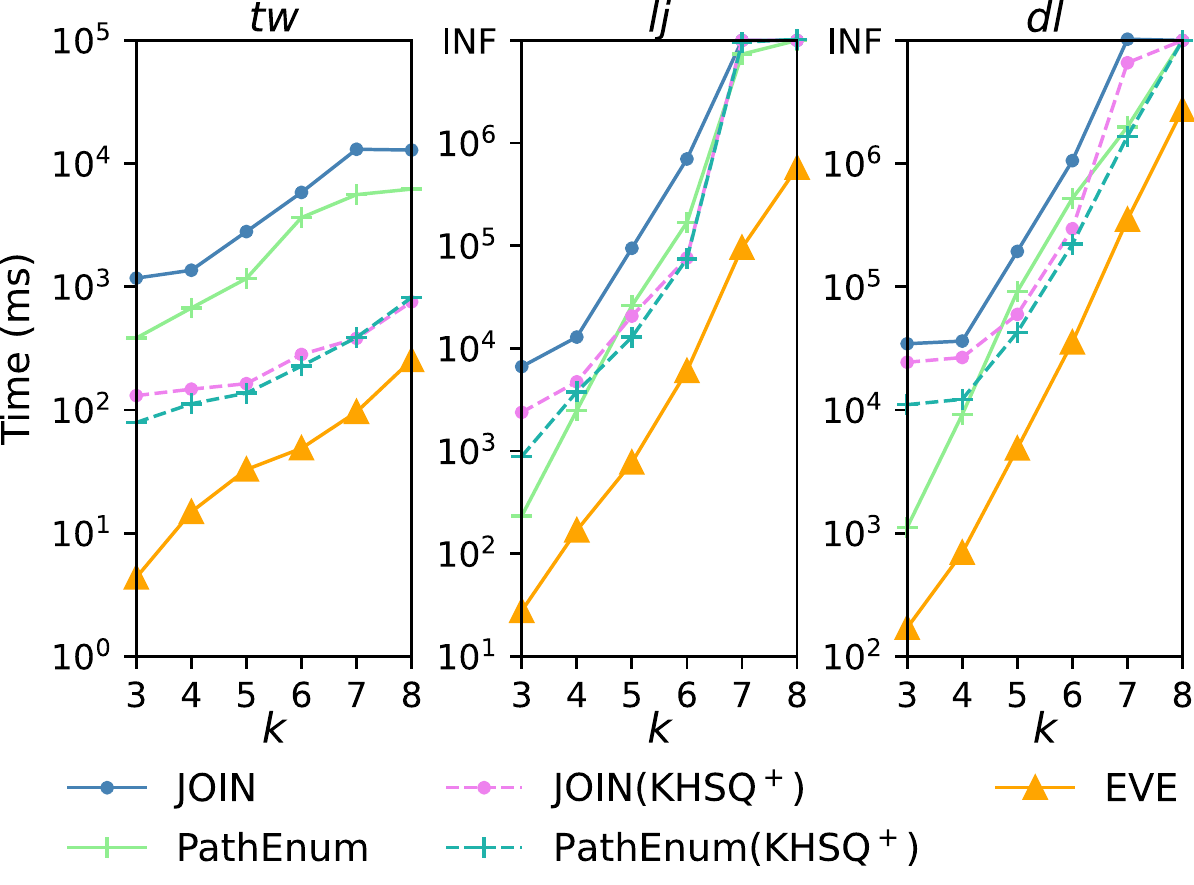}}
	\caption{
 Effect of $k$ on the average coverage ratio and time efficiency.
 }
    \label{fig: SPGratios-compare enhanced baselines}
\end{figure}

\textbf{Evaluating the coverage ratio}.
Since the number of edges in $SPG_k$ (denoted by $|E(SPG_k)|$) are bounded by $|E|$, we define the coverage ratio as $r_C=\frac{|E(SPG_k)|}{|E|}$ ($r_C \leq 1$). 
Figure~\ref{fig: SPGratios-compare enhanced baselines}(a) presents the average coverage ratio with respect to different $k$ among all graphs.
To make it clear, we split the results into two subfigures. We can see that  graphs with larger  average vertex degree tend to have higher coverage ratios, since larger degree indicates denser connection between query vertices $s$ and $t$ intuitively.

\textbf{Evaluating the redundant ratio}.
We compare the size of upper-bound graph $SPG_k^u$ with the answer graph $SPG_k$ for $k \geq 5$, since $SPG_k^u=SPG_k$ when $k\leq 4$. 
To measure the difference between the number of edges in $SPG_k^u$ and $SPG_k$, we define redundant ratio as $r_D=\frac{|E(SPG_k^u)|-|E(SPG_k)|}{|E(SPG_k)|}$ for each query.
Average $r_D$ for queries on each graph and $k$ are summarised in Table~\ref{tab:UPGdifferenceRatio}, where $0$ indicates that $SPG_k^u=SPG_k$. 
For most graphs except graphs \textit{st}, \textit{bs} and \textit{gg}, $SPG_k^u$ only has less than $0.05\%$ redundant edges.
Hence, $SPG_k^u$ computed in $\mathcal{O}(k^2|E|)$ is quite tight, reducing the search space sharply.

\begin{table}[t]
\caption{Speedups for hop-constrained \textit{s-t} path enumeration}
\label{fig:speedups}
\footnotesize
\begin{tabular}{c|c|cccccccccc}
\hline
                       & $k$ & \textit{ps} & \textit{sf} & \textit{bk} & \textit{tw} & \textit{bs} & \textit{wt}           & \textit{lj}           & \textit{dl}           & \textit{fr}           & \textit{hg}           \\ \hline
\multirow{4}{*}{KHSQ}  & 3 & 0.4  & 0.4  & 0.2  & 0.2  & 0.3  & \textless{}0.1 & \textless{}0.1 & \textless{}0.1 & \textless{}0.1 & <0.1              \\
                       & 4 & 0.8  & 0.2  & 0.1  & 0.1  & 0.2  & 0.1            & \textless{}0.1 & \textless{}0.1 & <0.1              & \textless{}0.1 \\
                       & 5 & 0.6  & 0.3  & 0.1  & 0.2  & 0.2  & 0.6            & \textless{}0.1 & \textless{}0.1 & <0.1              & <0.1              \\
                       & 6 & -    & 0.2  & 0.1  & 0.5  & 0.3  & 0.5            & \textless{}0.1 & 0.1              & <0.1              & -              \\ \hline
\multirow{4}{*}{KHSQ$^+$} & 3 & 0.7  & 1.6  & 1.5  & 3.7  & 1.3  & 0.3            & 0.2            & 0.1            & 0.2            & 0.2              \\
                       & 4 & 0.8  & 0.9  & 1.1  & 3.1  & 1.1  & 2.2            & 0.4            & 0.5            & 1.0            & 0.5              \\
                       & 5 & 0.6  & 2.1  & 1.9  & 11.8 & 1.2  & 1.4            & 2.5            & 1.6            & 3.2              & 1.2              \\
                       & 6 & -    & 1.5  & 1.9  & 23   & 0.7  & 0.8            & 3.6            & 2.3            & 1.7            & -       \\ \hline 
\multirow{4}{*}{EVE}   & 3 & 3.1  & 1.4  & 2.6  & 2.2  & 2.2  & 4.3            & 1.2            & 2.0            & 3.1            & 4.0            \\
                       & 4 & 1.3  & 1.9  & 2.1  & 3.4  & 1.8  & 15.2           & 4.1            & 7.4            & 23.1           & 6.6            \\
                       & 5 & 1.0  & 4.3  & 4.8  & 16.1 & 2.2  & 9.5            & 22.4           & 13.0             & 36.7           & 6.1            \\
                       & 6 & -    & 1.9  & 3.5  & 38.5 & 1.1  & 1.4            & 20.0             & 10.7           & 12             & -              \\ \hline
\end{tabular}
\end{table}

\subsection{Speedups for Simple Path Enumeration}
\label{sec:speedups}

We conduct experiments to illustrate that the simple path graph $SPG_k(s,t)$  helps to speed up hop-constrained $s$-$t$ path enumeration.
\textit{PathEnum} \cite{PathEnum} is the state-of-art algorithm for path enumeration.
For each query $\langle s,t,k\rangle$, we can first compute $SPG_k(s,t)$ by our proposed method \textit{EVE} and then use it as search space for \textit{PathEnum}, i.e., replacing the original graph $G$ with $SPG_k(s,t)$.
 
Speedups ($\frac{ \text{time of \textit{PathEnum} on the original graph } G }{\text{time of \textit{PathEnum} on } SPG_k(s,t) + \text{time of generating } SPG_k(s,t)}$) by varying datasets and $k$ are summarised in Table~\ref{fig:speedups}. 
For graphs \textit{tw}, \textit{wt}, \textit{lj}, \textit{dl}, and \textit{fr}, \textit{PathEnum} can be accelerated by up to an order of magnitude 
without any modification.

Moreover, the $k$-hop $s$-$t$ subgraph $G_{st}^k$ \cite{PathGraph} can also be used as search space for \textit{PathEnum},
which contains all $s$-$t$ paths within $k$ hops but those paths are not required to be simple paths.
However, generating $G_{st}^k$ with the state-of-the-art \textit{KHSQ} algorithm \cite{PathGraph} cannot accelerate \textit{PathEnum} (speedups $<1$), as shown in Table~\ref{fig:speedups}.
And we further develop its optimized algorithm \textit{KHSQ}$^+$, using adaptive bi-directional search (see Section~\ref{sec: adaptive bi-directional BFS}) instead of single directional BFS in \textit{KHSQ} for distance computation.
\textit{KHSQ$^+$} has much better performance than \textit{KHSQ}, but is still less efficient than $SPG_k(s,t)$ that is generated by \textit{EVE},
since $SPG_k(s,t)$ is a subgraph of $G_{st}^k$ and some time-consuming cycles contained in $G_{st}^k$ are avoided for \textit{PathEnum}.

\subsection{Generating Simple Path Graph based on $\mathbf{G_{st}^k}$}

For generating $SPG_k(s,t)$, we can first compute $G_{st}^k$ and then apply \textit{JOIN} \cite{JOIN2020, JOIN2021} (\textit{PathEnum} \cite{PathEnum}) on $G_{st}^k$.
As discussed in Section~\ref{sec:speedups}, \textit{KHSQ$^+$} is used instead of \textit{KHSQ} \cite{PathGraph} for efficiently computing $G_{st}^k$.
In this way, the modified \textit{JOIN} (\textit{PathEnum}) for generating $SPG_k(s,t)$ is faster than naive baselines. 
Table~\ref{tab: speedups of baselines} reports speedups ($\frac{\text{Time cost of \textit{JOIN} (\textit{PathEnum}) on } G}{\text{Time cost of \textit{JOIN} (\textit{PathEnum}) on } G_{st}^k}$) for $k=6$ except those run out of time.
However, by varying $k$ they are still not comparable with the proposed \textit{EVE} method.
As shown in Figure~\ref{fig: SPGratios-compare enhanced baselines}(b), even for graphs \textit{tw}, \textit{lj}, and \textit{dl} where significant speedups are achieved in Table~\ref{tab: speedups of baselines}, \textit{EVE} is still far more efficient than baselines enhanced by \textit{KHSQ$^+$}.

\begin{table}[t]
\caption{Speedups for generating $\mathbf{SPG_k(s,t)}$ on $\mathbf{G^k_{st}}$ ($\mathbf{k=6}$)}
\label{tab: speedups of baselines}
\footnotesize
\begin{tabular}{c|ccccccccccc}
\hline
\textbf{Dataset}  & \textit{wn} & \textit{uk} & \textit{sf} & \textit{bk} & \textit{tw} & \textit{bs} & \textit{gg} & \textit{wt} & \textit{lj} & \textit{dl} & \textit{fr} \\
\hline
\textbf{JOIN}     & $1.3$       & $2.1$       & $1.4$       & $1.9$       & $20.7$      & $0.4$       & $0.8$       & $3.1$       & $9.2$       & $3.6$       & -           \\
\textbf{PathEnum} & $0.9$       & $2.5$       & $1.9$       & $1.3$       & $16.0$      & $0.5$       & $1.1$       & $0.9$       & $2.3$       & $2.4$       & $2.6$   \\ 
\hline
\end{tabular}
\end{table}

\subsection{Case Study}
\label{sec:case study}

It has been shown that simple path graph works for visualizing connections between vertices $s$ and $t$ in Figure~\ref{fig:relfinder-pathcount}. 
Next, we introduce how \textit{EVE} helps identify fraudulent activities in a transaction network with millions of vertices and edges from an e-commerce company.
For a transaction (edge) $e(t,s)$ at time $T_0$, the goal is to find vertices and edges involved in any $(k+1)$-hop constrained cycles through $e(t,s)$, where all transaction time is required to be within $\Delta T$ days (i.e., only consider edges at time $[T_0-\Delta T, T_0]$ when searching).
The generated $SPG_k(s,t)$ for $k=5$ and $\Delta T=7$ is shown in Figure~\ref{fig:case study-related work}(a), in which accounts (vertices) and transactions (edges) are suspicious since they form simple cycle(s) in a short time period.

\section{Related Work}
\label{sec:Related Works}

\begin{figure}[t]
  \centering
  \includegraphics[width=\linewidth]{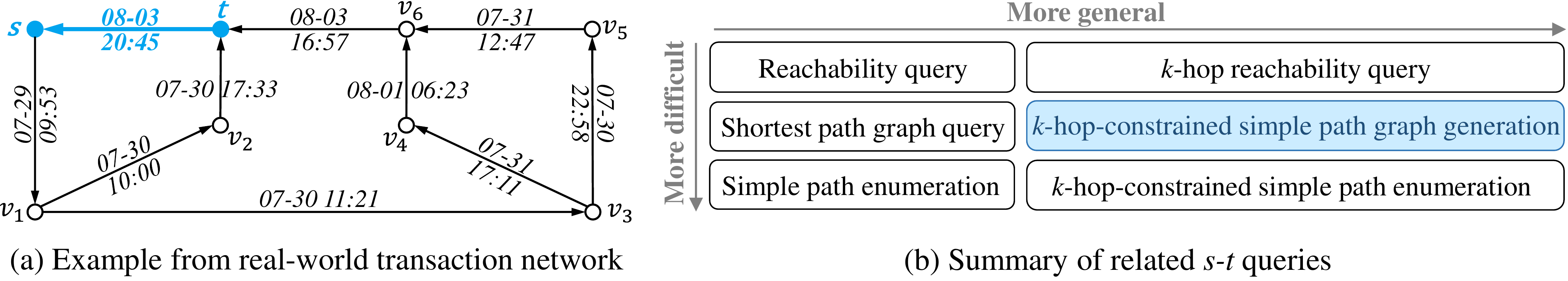}
  \caption{Case study and  summary of related work}
  \label{fig:case study-related work}
\end{figure}

In this section, we review related $s$-$t$ queries which are widely applied in the real world, as summarized in Figure~\ref{fig:case study-related work}(b).

\textbf{Reachability and $k$-hop reachability Query}.
Answering reachability queries is one of the fundamental graph queries and has been extensively studied \cite{BFL, PLL, TFlabel, IP, GRAIL, FELINE, DAGreduction}.
Since the number of hops indicates the level of influence $s$ has over $t$, many applications can benefit from $k$-hop reachability queries, which asks whether a vertex $s$ can reach $t$ within $k$ hops \cite{kReach2012, kReach2014, BFSI, HT, ESTI}.
$k$-hop reachability problem is more general since when $k=\infty$, we can obtain the answer to the reachability problem.

\textbf{Path Graph Queries}.
\textit{QbS} \cite{QbS} defines an $s$-$t$ shortest path graph \cite{QbS} that contains exactly all the shortest paths from $s$ to $t$.
It computes a sketch from the labeling scheme and then conducts an online search to return the exact answer.
However, the information provided by shortest path graph may be too limited, since paths slightly longer than shortest distance may also be of user interest.
It is not trivial to extend \textit{QbS} to our problem, since it is still inevitable for examining edges out of shortest path graph when $k$ is larger than shortest distance.
Liu et al. \cite{PathGraph} defines the $k$-hop $s$-$t$ subgraph query which returns the subgraph containing all paths from $s$ to $t$ within $k$ hops, but those paths are not required to be simple paths.
\textit{TransferPattern} \cite{TransferPattern} delivers a graph that just contains multi-criteria shortest paths by allowing multiple copies of vertices. 
It not only increases the size but also is unfriendly to display the underlying connections as it is not a subgraph of the input graph.

\textbf{(Hop-constrained) Simple Path Enumeration}.
There have been a bunch of works for simple path enumeration \cite{spe1, spe2, spe3, spe4}.
Hop-constrained $s$-$t$ simple path enumeration aims at finding all simple paths from $s$ to $t$ within $k$ hops, and recent works include \textit{TDFS} \cite{TDFS}, \textit{BC-DFS}, \textit{JOIN} \cite{JOIN2020, JOIN2021} and \textit{PathEnum} \cite{PathEnum}.
\textit{TDFS} \cite{TDFS} ensures that each vertex $v$ explored during forward DFS will produce at least an output path, which is achieved by backward BFS from target vertex $t$ at each step.
\textit{BC-DFS} \cite{JOIN2020, JOIN2021} is a barrier-pruning-based DFS algorithm, which prevents falling into the same trap twice by maintaining barriers during exploration. 
To improve query response time, \textit{JOIN} \cite{JOIN2020, JOIN2021} first finds some cutting vertices, then concatenates partial paths enumerated by \textit{BC-DFS}.
The time complexity of \textit{TDFS}, \textit{BC-DFS} and \textit{JOIN} are $\mathcal{O}(\delta k|E|)$, where $\delta$ is the number of output paths.
\textit{PathEnum} \cite{PathEnum} first builds a lightweight online index based on shortest distances, then uses a cost-based query optimizer to select DFS-based or join-based method for answering queries. 
As discussed before, enumerating all $k$-hop-constrained $s$-$t$ simple paths by above approaches is a straightforward solution for generating simple path graph $SPG_k(s,t)$. 
On the other hand, with the output $SPG_k(s,t)$ from our \textit{EVE} method, these approaches can also be accelerated by taking $SPG_k(s,t)$ as their search space.

\textbf{Pruning Techniques}. 
Forward-looking pruning adopts the general idea of checking distances \cite{TDFS, PathGraph}. 
In \cite{TDFS} distances are computed at each step of DFS, since vertices in DFS stack are removed to avoid vertex reuse for simple path.
In contrast, \cite{PathGraph} pre-computes all distances once but paths are not required to be simple. 
\textit{EVE} targets at the simple path graph, but also computes distances only once since essential vertices encode vertex reuse.
Moreover, though the idea of vertex reordering has been widely used with various goals \cite{QbS,PLL,BFSreordering,AstarCSP,LCR2020}, search ordering strategies are developed based on the tight upper-bound graph and use specialized keys (distance, $|In_D|$ and $|Out_A|$) for sorting neighbors to find valid paths earlier.

To prune spurious candidates when addressing the subgraph isomorphism problem, the algorithm \textit{KARPET} \cite{RankedEnumCQ,AnyK} adopts the idea of two-phase traversal similar to bi-directional propagation, but utilizes the query pattern to remove the vertices against the label constraints from the candidate graph in the first traversal.

\section{Conclusion}
\label{sec:Conclusions}
In this paper, we formalize the $k$-hop-constrained $s$-$t$ simple path graph generation problem, which has a wide range of applications.
We prove its NP-hardness on directed graphs and propose a method \textit{EVE}  to tackle this challenging problem. 
It does not need to enumerate all paths powered by the essential vertices. 
Moreover, a tight upper-bound graph is derived. To verify undetermined edges, a DFS-oriented search based on carefully designed orders is proposed.
Extensive experiments show that \textit{EVE} significantly outperforms all baselines, and it also helps to accelerate other graph queries such as hop-constrained simple path enumeration.

\begin{acks}
This work is supported in part by the National Natural Science Foundation of China under Grant U20B2046. Weiguo Zheng is the corresponding author.
\end{acks}

\bibliographystyle{ACM-Reference-Format}
\bibliography{references}


\begin{thebibliography}{48}


\ifx \showCODEN    \undefined \def \showCODEN     #1{\unskip}     \fi
\ifx \showDOI      \undefined \def \showDOI       #1{#1}\fi
\ifx \showISBNx    \undefined \def \showISBNx     #1{\unskip}     \fi
\ifx \showISBNxiii \undefined \def \showISBNxiii  #1{\unskip}     \fi
\ifx \showISSN     \undefined \def \showISSN      #1{\unskip}     \fi
\ifx \showLCCN     \undefined \def \showLCCN      #1{\unskip}     \fi
\ifx \shownote     \undefined \def \shownote      #1{#1}          \fi
\ifx \showarticletitle \undefined \def \showarticletitle #1{#1}   \fi
\ifx \showURL      \undefined \def \showURL       {\relax}        \fi
\providecommand\bibfield[2]{#2}
\providecommand\bibinfo[2]{#2}
\providecommand\natexlab[1]{#1}
\providecommand\showeprint[2][]{arXiv:#2}

\bibitem[Ahmadi et~al\mbox{.}(2021)]%
        {AstarCSP}
\bibfield{author}{\bibinfo{person}{Saman Ahmadi}, \bibinfo{person}{Guido Tack},
  \bibinfo{person}{Daniel~Damir Harabor}, {and} \bibinfo{person}{P. Kilby}.}
  \bibinfo{year}{2021}\natexlab{}.
\newblock \showarticletitle{A Fast Exact Algorithm for the Resource Constrained
  Shortest Path Problem}. In \bibinfo{booktitle}{\emph{AAAI}}.
\newblock


\bibitem[Ahuja et~al\mbox{.}(1993)]%
        {AdaptiveBidirectional_2}
\bibfield{author}{\bibinfo{person}{Ravindra~K. Ahuja},
  \bibinfo{person}{Thomas~L. Magnanti}, {and} \bibinfo{person}{James~B.
  Orlin}.} \bibinfo{year}{1993}\natexlab{}.
\newblock \showarticletitle{Network Flows: Theory, Algorithms, and
  Applications}.
\newblock


\bibitem[Alon et~al\mbox{.}(1995)]%
        {color-coding}
\bibfield{author}{\bibinfo{person}{Noga Alon}, \bibinfo{person}{Raphael
  Yuster}, {and} \bibinfo{person}{Uri Zwick}.} \bibinfo{year}{1995}\natexlab{}.
\newblock \showarticletitle{Color-coding}.
\newblock \bibinfo{journal}{\emph{Journal of the ACM (JACM)}}
  \bibinfo{volume}{42}, \bibinfo{number}{4} (\bibinfo{year}{1995}),
  \bibinfo{pages}{844--856}.
\newblock


\bibitem[Baier et~al\mbox{.}(2010)]%
        {MinCut}
\bibfield{author}{\bibinfo{person}{Georg Baier}, \bibinfo{person}{Thomas
  Erlebach}, \bibinfo{person}{Alexander Hall}, \bibinfo{person}{Ekkehard
  K{\"{o}}hler}, \bibinfo{person}{Petr Kolman}, \bibinfo{person}{Ondrej
  Pangr{\'{a}}c}, \bibinfo{person}{Heiko Schilling}, {and}
  \bibinfo{person}{Martin Skutella}.} \bibinfo{year}{2010}\natexlab{}.
\newblock \showarticletitle{Length-bounded cuts and flows}.
\newblock \bibinfo{journal}{\emph{{ACM} Trans. Algorithms}}
  \bibinfo{volume}{7}, \bibinfo{number}{1} (\bibinfo{year}{2010}),
  \bibinfo{pages}{4:1--4:27}.
\newblock


\bibitem[Bast et~al\mbox{.}(2010)]%
        {TransferPattern}
\bibfield{author}{\bibinfo{person}{Hannah Bast}, \bibinfo{person}{Erik
  Carlsson}, \bibinfo{person}{Arno Eigenwillig}, \bibinfo{person}{Robert
  Geisberger}, \bibinfo{person}{Chris Harrelson}, \bibinfo{person}{Veselin
  Raychev}, {and} \bibinfo{person}{Fabien Viger}.}
  \bibinfo{year}{2010}\natexlab{}.
\newblock \showarticletitle{Fast Routing in Very Large Public Transportation
  Networks Using Transfer Patterns}. In \bibinfo{booktitle}{\emph{Algorithms --
  ESA 2010}}. \bibinfo{pages}{290--301}.
\newblock


\bibitem[B{\"{a}}umer et~al\mbox{.}(2014)]%
        {SciVis}
\bibfield{author}{\bibinfo{person}{Frederik~Simon B{\"{a}}umer},
  \bibinfo{person}{Jangwon Gim}, \bibinfo{person}{Do{-}Heon Jeong},
  \bibinfo{person}{Michaela Geierhos}, {and} \bibinfo{person}{Hanmin Jung}.}
  \bibinfo{year}{2014}\natexlab{}.
\newblock \showarticletitle{Linked Open Data System for Scientific Data Sets}.
  In \bibinfo{booktitle}{\emph{IPaMin 2014}} \emph{(\bibinfo{series}{{CEUR}
  Workshop Proceedings}, Vol.~\bibinfo{volume}{1292})}.
\newblock


\bibitem[Birmel{\'{e}} et~al\mbox{.}(2013)]%
        {spe3}
\bibfield{author}{\bibinfo{person}{Etienne Birmel{\'{e}}},
  \bibinfo{person}{Rui~A. Ferreira}, \bibinfo{person}{Roberto Grossi},
  \bibinfo{person}{Andrea Marino}, \bibinfo{person}{Nadia Pisanti},
  \bibinfo{person}{Romeo Rizzi}, {and} \bibinfo{person}{Gustavo Sacomoto}.}
  \bibinfo{year}{2013}\natexlab{}.
\newblock \showarticletitle{Optimal Listing of Cycles and st-Paths in
  Undirected Graphs}. In \bibinfo{booktitle}{\emph{{SODA} 2013}}.
  \bibinfo{pages}{1884--1896}.
\newblock


\bibitem[Cabrera et~al\mbox{.}(2020)]%
        {ConstrainedShortestPath}
\bibfield{author}{\bibinfo{person}{Nicol{\'{a}}s Cabrera},
  \bibinfo{person}{Andr{\'{e}}s~L. Medaglia}, \bibinfo{person}{Leonardo
  Lozano}, {and} \bibinfo{person}{Daniel Duque}.}
  \bibinfo{year}{2020}\natexlab{}.
\newblock \showarticletitle{An exact bidirectional pulse algorithm for the
  constrained shortest path}.
\newblock \bibinfo{journal}{\emph{Networks}} \bibinfo{volume}{76},
  \bibinfo{number}{2} (\bibinfo{year}{2020}), \bibinfo{pages}{128--146}.
\newblock


\bibitem[Cai and Zheng(2021)]%
        {ESTI}
\bibfield{author}{\bibinfo{person}{Yuzheng Cai} {and} \bibinfo{person}{Weiguo
  Zheng}.} \bibinfo{year}{2021}\natexlab{}.
\newblock \showarticletitle{{ESTI:} Efficient k-Hop Reachability Querying over
  Large General Directed Graphs}. In \bibinfo{booktitle}{\emph{{DASFAA} 2021
  International Workshops - BDQM, GDMA, MLDLDSA, MobiSocial, and MUST}}.
  \bibinfo{publisher}{Springer}, \bibinfo{pages}{71--89}.
\newblock


\bibitem[Cheng et~al\mbox{.}(2013)]%
        {TFlabel}
\bibfield{author}{\bibinfo{person}{James Cheng}, \bibinfo{person}{Silu Huang},
  \bibinfo{person}{Huanhuan Wu}, {and} \bibinfo{person}{Ada~Wai{-}Chee Fu}.}
  \bibinfo{year}{2013}\natexlab{}.
\newblock \showarticletitle{TF-Label: a topological-folding labeling scheme for
  reachability querying in a large graph}. In
  \bibinfo{booktitle}{\emph{{SIGMOD} 2013}}. \bibinfo{pages}{193--204}.
\newblock


\bibitem[Cheng et~al\mbox{.}(2012)]%
        {kReach2012}
\bibfield{author}{\bibinfo{person}{James Cheng}, \bibinfo{person}{Zechao
  Shang}, \bibinfo{person}{Hong Cheng}, \bibinfo{person}{Haixun Wang}, {and}
  \bibinfo{person}{Jeffrey~Xu Yu}.} \bibinfo{year}{2012}\natexlab{}.
\newblock \showarticletitle{K-Reach: Who is in Your Small World}.
\newblock \bibinfo{journal}{\emph{Proc. {VLDB} Endow.}} \bibinfo{volume}{5},
  \bibinfo{number}{11} (\bibinfo{year}{2012}), \bibinfo{pages}{1292--1303}.
\newblock


\bibitem[Cheng et~al\mbox{.}(2014)]%
        {kReach2014}
\bibfield{author}{\bibinfo{person}{James Cheng}, \bibinfo{person}{Zechao
  Shang}, \bibinfo{person}{Hong Cheng}, \bibinfo{person}{Haixun Wang}, {and}
  \bibinfo{person}{Jeffrey~Xu Yu}.} \bibinfo{year}{2014}\natexlab{}.
\newblock \showarticletitle{Efficient processing of k-hop reachability
  queries}.
\newblock \bibinfo{journal}{\emph{{VLDB} J.}} \bibinfo{volume}{23},
  \bibinfo{number}{2} (\bibinfo{year}{2014}), \bibinfo{pages}{227--252}.
\newblock


\bibitem[Du et~al\mbox{.}(2019)]%
        {HT}
\bibfield{author}{\bibinfo{person}{Ming Du}, \bibinfo{person}{Anping Yang},
  \bibinfo{person}{Junfeng Zhou}, \bibinfo{person}{Xian Tang},
  \bibinfo{person}{Ziyang Chen}, {and} \bibinfo{person}{Yanfei Zuo}.}
  \bibinfo{year}{2019}\natexlab{}.
\newblock \showarticletitle{\emph{HT}: {A} Novel Labeling Scheme for k-Hop
  Reachability Queries on DAGs}.
\newblock \bibinfo{journal}{\emph{{IEEE} Access}}  \bibinfo{volume}{7}
  (\bibinfo{year}{2019}), \bibinfo{pages}{172110--172122}.
\newblock


\bibitem[Fomin et~al\mbox{.}(2018)]%
        {directeds-t}
\bibfield{author}{\bibinfo{person}{Fedor~V Fomin}, \bibinfo{person}{Daniel
  Lokshtanov}, \bibinfo{person}{Fahad Panolan}, \bibinfo{person}{Saket
  Saurabh}, {and} \bibinfo{person}{Meirav Zehavi}.}
  \bibinfo{year}{2018}\natexlab{}.
\newblock \showarticletitle{Long directed (s, t)-path: FPT algorithm}.
\newblock \bibinfo{journal}{\emph{Inform. Process. Lett.}}
  \bibinfo{volume}{140} (\bibinfo{year}{2018}), \bibinfo{pages}{8--12}.
\newblock


\bibitem[Fortune et~al\mbox{.}(1980)]%
        {ProofNP}
\bibfield{author}{\bibinfo{person}{Steven Fortune}, \bibinfo{person}{John~E.
  Hopcroft}, {and} \bibinfo{person}{James Wyllie}.}
  \bibinfo{year}{1980}\natexlab{}.
\newblock \showarticletitle{The Directed Subgraph Homeomorphism Problem}.
\newblock \bibinfo{journal}{\emph{Theor. Comput. Sci.}}  \bibinfo{volume}{10}
  (\bibinfo{year}{1980}), \bibinfo{pages}{111--121}.
\newblock


\bibitem[Garc{\'{\i}}a{-}Godoy et~al\mbox{.}(2011)]%
        {BioVis}
\bibfield{author}{\bibinfo{person}{Mar{\'{\i}}a~Jes{\'{u}}s
  Garc{\'{\i}}a{-}Godoy}, \bibinfo{person}{Ismael~Navas Delgado}, {and}
  \bibinfo{person}{Jos{\'{e}} Francisco~Aldana Montes}.}
  \bibinfo{year}{2011}\natexlab{}.
\newblock \showarticletitle{Bioqueries: a social community sharing experiences
  while querying biological linked data}. In
  \bibinfo{booktitle}{\emph{{SWAT4LS} 2011}}. \bibinfo{pages}{24--31}.
\newblock


\bibitem[Goldberg and Harrelson(2005)]%
        {bidirectional}
\bibfield{author}{\bibinfo{person}{Andrew~V. Goldberg} {and}
  \bibinfo{person}{Chris Harrelson}.} \bibinfo{year}{2005}\natexlab{}.
\newblock \showarticletitle{Computing the shortest path: A search meets graph
  theory}.
\newblock \bibinfo{journal}{\emph{symposium on discrete algorithms}}
  (\bibinfo{year}{2005}).
\newblock


\bibitem[Grossi(2016)]%
        {spe1}
\bibfield{author}{\bibinfo{person}{Roberto Grossi}.}
  \bibinfo{year}{2016}\natexlab{}.
\newblock \showarticletitle{Enumeration of Paths, Cycles, and Spanning Trees}.
\newblock In \bibinfo{booktitle}{\emph{Encyclopedia of Algorithms}}.
  \bibinfo{pages}{640--645}.
\newblock


\bibitem[Grossi et~al\mbox{.}(2018)]%
        {spe4}
\bibfield{author}{\bibinfo{person}{Roberto Grossi}, \bibinfo{person}{Andrea
  Marino}, {and} \bibinfo{person}{Luca Versari}.}
  \bibinfo{year}{2018}\natexlab{}.
\newblock \showarticletitle{Efficient Algorithms for Listing k Disjoint
  st-Paths in Graphs}. In \bibinfo{booktitle}{\emph{{LATIN} 2018}},
  Vol.~\bibinfo{volume}{10807}. \bibinfo{pages}{544--557}.
\newblock


\bibitem[Heim et~al\mbox{.}(2009)]%
        {RelFinder2009}
\bibfield{author}{\bibinfo{person}{Philipp Heim}, \bibinfo{person}{Sebastian
  Hellmann}, \bibinfo{person}{Jens Lehmann}, \bibinfo{person}{Steffen Lohmann},
  {and} \bibinfo{person}{Timo Stegemann}.} \bibinfo{year}{2009}\natexlab{}.
\newblock \showarticletitle{RelFinder: Revealing Relationships in {RDF}
  Knowledge Bases}. In \bibinfo{booktitle}{\emph{{SAMT} 2009}},
  Vol.~\bibinfo{volume}{5887}. \bibinfo{pages}{182--187}.
\newblock


\bibitem[Holzer et~al\mbox{.}(2005)]%
        {AdaptiveBidirectional_1}
\bibfield{author}{\bibinfo{person}{Martin Holzer}, \bibinfo{person}{Frank
  Schulz}, \bibinfo{person}{Dorothea Wagner}, {and} \bibinfo{person}{Thomas
  Willhalm}.} \bibinfo{year}{2005}\natexlab{}.
\newblock \showarticletitle{Combining speed-up techniques for shortest-path
  computations}.
\newblock \bibinfo{journal}{\emph{ACM Journal of Experimental Algorithms}}
  (\bibinfo{year}{2005}).
\newblock


\bibitem[Kunegis(2013)]%
        {konect}
\bibfield{author}{\bibinfo{person}{J{\'{e}}r{\^{o}}me Kunegis}.}
  \bibinfo{year}{2013}\natexlab{}.
\newblock \showarticletitle{{KONECT:} the Koblenz network collection}. In
  \bibinfo{booktitle}{\emph{{WWW} '13}}. \bibinfo{pages}{1343--1350}.
\newblock


\bibitem[Leskovec et~al\mbox{.}(2010)]%
        {Signed-Network}
\bibfield{author}{\bibinfo{person}{Jure Leskovec}, \bibinfo{person}{Daniel
  Huttenlocher}, {and} \bibinfo{person}{Jon Kleinberg}.}
  \bibinfo{year}{2010}\natexlab{}.
\newblock \showarticletitle{Signed networks in social media}. In
  \bibinfo{booktitle}{\emph{Proceedings of the SIGCHI conference on human
  factors in computing systems}}. \bibinfo{pages}{1361--1370}.
\newblock


\bibitem[Leskovec and Krevl(2014)]%
        {snapnets}
\bibfield{author}{\bibinfo{person}{Jure Leskovec} {and} \bibinfo{person}{Andrej
  Krevl}.} \bibinfo{year}{2014}\natexlab{}.
\newblock \bibinfo{title}{{SNAP Datasets}: {Stanford} Large Network Dataset
  Collection}.
\newblock
\newblock


\bibitem[Liu et~al\mbox{.}(2021)]%
        {PathGraph}
\bibfield{author}{\bibinfo{person}{Yu Liu}, \bibinfo{person}{Qian Ge},
  \bibinfo{person}{Yue Pang}, {and} \bibinfo{person}{Lei Zou}.}
  \bibinfo{year}{2021}\natexlab{}.
\newblock \showarticletitle{Hop-Constrained Subgraph Query and Summarization on
  Large Graphs}. In \bibinfo{booktitle}{\emph{{DASFAA} 2021 International
  Workshops - BDQM, GDMA, MLDLDSA, MobiSocial, and MUST}},
  Vol.~\bibinfo{volume}{12680}. \bibinfo{pages}{123--139}.
\newblock


\bibitem[Lohmann et~al\mbox{.}(2010)]%
        {RelFinder2010}
\bibfield{author}{\bibinfo{person}{Steffen Lohmann}, \bibinfo{person}{Philipp
  Heim}, \bibinfo{person}{Timo Stegemann}, {and} \bibinfo{person}{J{\"{u}}rgen
  Ziegler}.} \bibinfo{year}{2010}\natexlab{}.
\newblock \showarticletitle{The RelFinder user interface: interactive
  exploration of relationships between objects of interest}. In
  \bibinfo{booktitle}{\emph{{IUI} 2010}}. \bibinfo{pages}{421--422}.
\newblock


\bibitem[Peng et~al\mbox{.}(2021)]%
        {JOIN2021}
\bibfield{author}{\bibinfo{person}{You Peng}, \bibinfo{person}{Xuemin Lin},
  \bibinfo{person}{Ying Zhang}, \bibinfo{person}{Wenjie Zhang},
  \bibinfo{person}{Lu Qin}, {and} \bibinfo{person}{Jingren Zhou}.}
  \bibinfo{year}{2021}\natexlab{}.
\newblock \showarticletitle{Efficient Hop-constrained s-t Simple Path
  Enumeration}.
\newblock \bibinfo{journal}{\emph{{VLDB} J.}} \bibinfo{volume}{30},
  \bibinfo{number}{5} (\bibinfo{year}{2021}), \bibinfo{pages}{799--823}.
\newblock


\bibitem[Peng et~al\mbox{.}(2020)]%
        {LCR2020}
\bibfield{author}{\bibinfo{person}{You Peng}, \bibinfo{person}{Ying Zhang},
  \bibinfo{person}{Xuemin Lin}, \bibinfo{person}{Lu Qin}, {and}
  \bibinfo{person}{Wenjie Zhang}.} \bibinfo{year}{2020}\natexlab{}.
\newblock \showarticletitle{Answering Billion-Scale Label-Constrained
  Reachability Queries within Microsecond}.
\newblock \bibinfo{journal}{\emph{Proc. VLDB Endow.}} \bibinfo{volume}{13},
  \bibinfo{number}{6} (\bibinfo{year}{2020}), \bibinfo{pages}{812–825}.
\newblock
\showISSN{2150-8097}


\bibitem[Peng et~al\mbox{.}(2019)]%
        {JOIN2020}
\bibfield{author}{\bibinfo{person}{You Peng}, \bibinfo{person}{Ying Zhang},
  \bibinfo{person}{Xuemin Lin}, \bibinfo{person}{Wenjie Zhang},
  \bibinfo{person}{Lu Qin}, {and} \bibinfo{person}{Jingren Zhou}.}
  \bibinfo{year}{2019}\natexlab{}.
\newblock \showarticletitle{Hop-constrained s-t Simple Path Enumeration:
  Towards Bridging Theory and Practice}.
\newblock \bibinfo{journal}{\emph{Proc. {VLDB} Endow.}} \bibinfo{volume}{13},
  \bibinfo{number}{4} (\bibinfo{year}{2019}), \bibinfo{pages}{463--476}.
\newblock


\bibitem[Qiu et~al\mbox{.}(2018)]%
        {Alibaba}
\bibfield{author}{\bibinfo{person}{Xiafei Qiu}, \bibinfo{person}{Wubin Cen},
  \bibinfo{person}{Zhengping Qian}, \bibinfo{person}{You Peng},
  \bibinfo{person}{Ying Zhang}, \bibinfo{person}{Xuemin Lin}, {and}
  \bibinfo{person}{Jingren Zhou}.} \bibinfo{year}{2018}\natexlab{}.
\newblock \showarticletitle{Real-time Constrained Cycle Detection in Large
  Dynamic Graphs}.
\newblock \bibinfo{journal}{\emph{Proc. {VLDB} Endow.}} \bibinfo{volume}{11},
  \bibinfo{number}{12} (\bibinfo{year}{2018}), \bibinfo{pages}{1876--1888}.
\newblock


\bibitem[Rizzi et~al\mbox{.}(2014)]%
        {TDFS}
\bibfield{author}{\bibinfo{person}{Romeo Rizzi}, \bibinfo{person}{Gustavo
  Sacomoto}, {and} \bibinfo{person}{Marie{-}France Sagot}.}
  \bibinfo{year}{2014}\natexlab{}.
\newblock \showarticletitle{Efficiently Listing Bounded Length st-Paths}. In
  \bibinfo{booktitle}{\emph{{IWOCA} 2014}}, Vol.~\bibinfo{volume}{8986}.
  \bibinfo{pages}{318--329}.
\newblock


\bibitem[Rossi and Ahmed(2015)]%
        {NetworkRepository}
\bibfield{author}{\bibinfo{person}{Ryan~A. Rossi} {and}
  \bibinfo{person}{Nesreen~K. Ahmed}.} \bibinfo{year}{2015}\natexlab{}.
\newblock \showarticletitle{The Network Data Repository with Interactive Graph
  Analytics and Visualization}. In \bibinfo{booktitle}{\emph{{AAAI} 2015}}.
  \bibinfo{pages}{4292--4293}.
\newblock


\bibitem[Sobrinho and Ferreira(2020)]%
        {ShortestWidestRouting}
\bibfield{author}{\bibinfo{person}{Jo{\~{a}}o~Luis Sobrinho} {and}
  \bibinfo{person}{Miguel~Alves Ferreira}.} \bibinfo{year}{2020}\natexlab{}.
\newblock \showarticletitle{Routing on Multiple Optimality Criteria}. In
  \bibinfo{booktitle}{\emph{{SIGCOMM} '20}}. \bibinfo{pages}{211--225}.
\newblock


\bibitem[Su et~al\mbox{.}(2017)]%
        {BFL}
\bibfield{author}{\bibinfo{person}{Jiao Su}, \bibinfo{person}{Qing Zhu},
  \bibinfo{person}{Hao Wei}, {and} \bibinfo{person}{Jeffrey~Xu Yu}.}
  \bibinfo{year}{2017}\natexlab{}.
\newblock \showarticletitle{Reachability Querying: Can It Be Even Faster?}
\newblock \bibinfo{journal}{\emph{TKDE}} \bibinfo{volume}{29},
  \bibinfo{number}{3} (\bibinfo{year}{2017}), \bibinfo{pages}{683--697}.
\newblock


\bibitem[Sun et~al\mbox{.}(2021)]%
        {PathEnum}
\bibfield{author}{\bibinfo{person}{Shixuan Sun}, \bibinfo{person}{Yuhang Chen},
  \bibinfo{person}{Bingsheng He}, {and} \bibinfo{person}{Bryan Hooi}.}
  \bibinfo{year}{2021}\natexlab{}.
\newblock \showarticletitle{PathEnum: Towards Real-Time Hop-Constrained s-t
  Path Enumeration}. In \bibinfo{booktitle}{\emph{{SIGMOD} '21}}.
  \bibinfo{pages}{1758--1770}.
\newblock


\bibitem[Tziavelis et~al\mbox{.}(2020)]%
        {RankedEnumCQ}
\bibfield{author}{\bibinfo{person}{Nikolaos Tziavelis}, \bibinfo{person}{Deepak
  Ajwani}, \bibinfo{person}{Wolfgang Gatterbauer}, \bibinfo{person}{Mirek
  Riedewald}, {and} \bibinfo{person}{Xiaofeng Yang}.}
  \bibinfo{year}{2020}\natexlab{}.
\newblock \showarticletitle{Optimal Algorithms for Ranked Enumeration of
  Answers to Full Conjunctive Queries}.
\newblock \bibinfo{journal}{\emph{Proc. VLDB Endow.}} \bibinfo{volume}{13},
  \bibinfo{number}{9} (\bibinfo{year}{2020}), \bibinfo{pages}{1582–1597}.
\newblock


\bibitem[Ueno et~al\mbox{.}(2017)]%
        {BFSreordering}
\bibfield{author}{\bibinfo{person}{Koji Ueno}, \bibinfo{person}{Toyotaro
  Suzumura}, \bibinfo{person}{Naoya Maruyama}, \bibinfo{person}{Katsuki
  Fujisawa}, {and} \bibinfo{person}{Satoshi Matsuoka}.}
  \bibinfo{year}{2017}\natexlab{}.
\newblock \showarticletitle{Efficient Breadth-First Search on Massively
  Parallel and Distributed-Memory Machines}.
\newblock \bibinfo{journal}{\emph{Data Science and Engineering}}
  (\bibinfo{year}{2017}).
\newblock


\bibitem[Veloso et~al\mbox{.}(2014)]%
        {FELINE}
\bibfield{author}{\bibinfo{person}{Ren{\^{e}}~Rodrigues Veloso},
  \bibinfo{person}{Lo{\"{\i}}c Cerf}, \bibinfo{person}{Wagner~Meira Jr.}, {and}
  \bibinfo{person}{Mohammed~J. Zaki}.} \bibinfo{year}{2014}\natexlab{}.
\newblock \showarticletitle{Reachability Queries in Very Large Graphs: {A} Fast
  Refined Online Search Approach}. In \bibinfo{booktitle}{\emph{{EDBT} 2014}}.
  \bibinfo{pages}{511--522}.
\newblock


\bibitem[Wang et~al\mbox{.}(2021)]%
        {QbS}
\bibfield{author}{\bibinfo{person}{Ye Wang}, \bibinfo{person}{Qing Wang},
  \bibinfo{person}{Henning Koehler}, {and} \bibinfo{person}{Yu Lin}.}
  \bibinfo{year}{2021}\natexlab{}.
\newblock \showarticletitle{Query-by-Sketch: Scaling Shortest Path Graph
  Queries on Very Large Networks}. In \bibinfo{booktitle}{\emph{{SIGMOD} '21}}.
  \bibinfo{pages}{1946--1958}.
\newblock


\bibitem[Wang and Crowcroft(1996)]%
        {routing}
\bibfield{author}{\bibinfo{person}{Zheng Wang} {and} \bibinfo{person}{Jon
  Crowcroft}.} \bibinfo{year}{1996}\natexlab{}.
\newblock \showarticletitle{Quality-of-Service Routing for Supporting
  Multimedia Applications}.
\newblock \bibinfo{journal}{\emph{{IEEE} J. Sel. Areas Commun.}}
  \bibinfo{volume}{14}, \bibinfo{number}{7} (\bibinfo{year}{1996}),
  \bibinfo{pages}{1228--1234}.
\newblock


\bibitem[Wei et~al\mbox{.}(2018)]%
        {IP}
\bibfield{author}{\bibinfo{person}{Hao Wei}, \bibinfo{person}{Jeffrey~Xu Yu},
  \bibinfo{person}{Can Lu}, {and} \bibinfo{person}{Ruoming Jin}.}
  \bibinfo{year}{2018}\natexlab{}.
\newblock \showarticletitle{Reachability querying: an independent permutation
  labeling approach}.
\newblock \bibinfo{journal}{\emph{{VLDB} J.}} \bibinfo{volume}{27},
  \bibinfo{number}{1} (\bibinfo{year}{2018}), \bibinfo{pages}{1--26}.
\newblock


\bibitem[Xie et~al\mbox{.}(2017)]%
        {BFSI}
\bibfield{author}{\bibinfo{person}{Xia Xie}, \bibinfo{person}{Xiaodong Yang},
  \bibinfo{person}{Xiaokang Wang}, \bibinfo{person}{Hai Jin},
  \bibinfo{person}{Duoqiang Wang}, {and} \bibinfo{person}{Xijiang Ke}.}
  \bibinfo{year}{2017}\natexlab{}.
\newblock \showarticletitle{{BFSI-B:} An improved K-hop graph reachability
  queries for cyber-physical systems}.
\newblock \bibinfo{journal}{\emph{Inf. Fusion}}  \bibinfo{volume}{38}
  (\bibinfo{year}{2017}), \bibinfo{pages}{35--42}.
\newblock


\bibitem[Yang et~al\mbox{.}(2015)]%
        {spe2}
\bibfield{author}{\bibinfo{person}{Runtao Yang}, \bibinfo{person}{Rui Gao},
  {and} \bibinfo{person}{Chengjin Zhang}.} \bibinfo{year}{2015}\natexlab{}.
\newblock \showarticletitle{A new algebraic approach to finding all simple
  paths and cycles in undirected graphs}. In \bibinfo{booktitle}{\emph{{ICIA}
  2015}}. \bibinfo{pages}{1887--1892}.
\newblock


\bibitem[Yang et~al\mbox{.}(2018)]%
        {AnyK}
\bibfield{author}{\bibinfo{person}{Xiaofeng Yang}, \bibinfo{person}{Deepak
  Ajwani}, \bibinfo{person}{Wolfgang Gatterbauer}, \bibinfo{person}{Patrick~K.
  Nicholson}, \bibinfo{person}{Mirek Riedewald}, {and}
  \bibinfo{person}{Alessandra Sala}.} \bibinfo{year}{2018}\natexlab{}.
\newblock \showarticletitle{Any-k: Anytime Top-k Tree Pattern Retrieval in
  Labeled Graphs} \emph{(\bibinfo{series}{WWW '18})}.
  \bibinfo{pages}{489–498}.
\newblock


\bibitem[Yano et~al\mbox{.}(2013)]%
        {PLL}
\bibfield{author}{\bibinfo{person}{Yosuke Yano}, \bibinfo{person}{Takuya
  Akiba}, \bibinfo{person}{Yoichi Iwata}, {and} \bibinfo{person}{Yuichi
  Yoshida}.} \bibinfo{year}{2013}\natexlab{}.
\newblock \showarticletitle{Fast and scalable reachability queries on graphs by
  pruned labeling with landmarks and paths}. In
  \bibinfo{booktitle}{\emph{CIKM'13}}. \bibinfo{pages}{1601--1606}.
\newblock


\bibitem[Yeh et~al\mbox{.}(2012)]%
        {PathWay}
\bibfield{author}{\bibinfo{person}{Cheng-Yu Yeh}, \bibinfo{person}{Hsiang-Yuan
  Yeh}, \bibinfo{person}{Carlos~Roberto Arias}, {and} \bibinfo{person}{Von-Wun
  Soo}.} \bibinfo{year}{2012}\natexlab{}.
\newblock \showarticletitle{Pathway detection from protein interaction networks
  and gene expression data using color-coding methods and A* search
  algorithms}.
\newblock \bibinfo{journal}{\emph{The Scientific World Journal}}
  \bibinfo{volume}{2012} (\bibinfo{year}{2012}).
\newblock


\bibitem[Yildirim et~al\mbox{.}(2012)]%
        {GRAIL}
\bibfield{author}{\bibinfo{person}{Hilmi Yildirim}, \bibinfo{person}{Vineet
  Chaoji}, {and} \bibinfo{person}{Mohammed~J. Zaki}.}
  \bibinfo{year}{2012}\natexlab{}.
\newblock \showarticletitle{{GRAIL:} a scalable index for reachability queries
  in very large graphs}.
\newblock \bibinfo{journal}{\emph{{VLDB} J.}} \bibinfo{volume}{21},
  \bibinfo{number}{4} (\bibinfo{year}{2012}), \bibinfo{pages}{509--534}.
\newblock


\bibitem[Zhou et~al\mbox{.}(2018)]%
        {DAGreduction}
\bibfield{author}{\bibinfo{person}{Junfeng Zhou}, \bibinfo{person}{Jeffrey~Xu
  Yu}, \bibinfo{person}{Na Li}, \bibinfo{person}{Hao Wei},
  \bibinfo{person}{Ziyang Chen}, {and} \bibinfo{person}{Xian Tang}.}
  \bibinfo{year}{2018}\natexlab{}.
\newblock \showarticletitle{Accelerating reachability query processing based on
  {DAG} reduction}.
\newblock \bibinfo{journal}{\emph{{VLDB} J.}} \bibinfo{volume}{27},
  \bibinfo{number}{2} (\bibinfo{year}{2018}), \bibinfo{pages}{271--296}.
\newblock


\end{thebibliography}


\end{document}